%% file: main.tex
\documentclass[twocolumn]{main}
\usepackage{graphicx}	
\usepackage{amsmath}	
\usepackage{amssymb}	
\usepackage{booktabs}

\usepackage{comment}

\shorttitle{SGR\,1935+2154 H.E.S.S. follow-up}
\shortauthors{H.E.S.S. collaboration}

\graphicspath{{./}{Figures/}}

\begin{document}

\title{Searching for TeV gamma-ray emission from SGR\,1935+2154 during its 2020 X-ray and radio bursting phase}

\correspondingauthor{H.~Ashkar, D. Kostunin, G. Rowell and F. Schüssler}
\email{contact.hess@hess-experiment.eu}
\input{authors_apj}

\begin{abstract}
Magnetar hyperflares are the most plausible explanation for fast radio bursts (FRB) --- enigmatic powerful radio pulses with durations of several milliseconds and high brightness temperatures. The first observational evidence for this scenario was obtained in 2020 April when a FRB was detected from the direction of the Galactic magnetar and soft gamma-ray repeater SGR\,1935+2154. The FRB was preceded by two gamma-ray outburst alerts by the BAT instrument aboard the Swift satellite, which triggered follow-up observations by the High Energy Stereoscopic System (H.E.S.S.). H.E.S.S. has observed SGR\,1935+2154 for 2 hr on 2020 April 28. The observations are coincident with X-ray bursts from the magnetar detected by INTEGRAL and Fermi-GBM, thus providing the first very high energy (VHE) gamma-ray observations of a magnetar in a flaring state. 
High-quality data acquired during these follow-up observations allow us to perform a search for short-time transients. No significant signal at energies $E>0.6$~TeV is found and upper limits on the persistent and transient emission are derived.
We here present the analysis of these observations and discuss the obtained results and prospects of the H.E.S.S. follow-up program for soft gamma-ray repeaters.
\end{abstract}

\keywords{Soft gamma-ray repeater -- Fast Radio Burst -- H.E.S.S. -- VHE gamma-rays}

\section{Introduction}



Soft gamma-ray repeaters (SGR) and anomalous X-ray pulsars (AXPs) are associated with highly magnetized neutron stars or magnetars. They generate bursts of emission at irregular time intervals. The crust of the neutron star is thought to break owing to the intense shifts of the ultrastrong magnetic field causing the emission of hard X-rays and gamma rays. During these short ($\sim$0.1\,s) bursts, the brightness of these objects can increase by a factor of 1000 or more. Within this category, the most extreme giant flares are so intense that in the case of the 2004 27 December event from SGR\,1806-20, they can influence the Earth's ionosphere and magnetic field  \citep[e.g.,][]{Inan:2007}. 

\subsection{SGRs and Nonthermal emission}
For many years it was unclear whether nonthermal emission mechanisms are involved in these burst activities. Recently the Large Area Telescope (LAT) aboard the Fermi satellite detected GeV gamma-ray emission from an extragalactic magnetar in the Sculptor galaxy group \citep{FermiMagnetar:2021}. The location of the recorded burst is consistent with several galaxies, including the starburst galaxy NGC\,253 (the Sculptor galaxy) at 3.5\,Mpc distance. The unusually long delay (19\,s) of the GeV emission from the first spike and quasi-periodic oscillation seen in the MeV signal suggest a giant magnetar flare causing gamma emission well outside the magnetar's light cylinder radius. At present, the GeV radiation is attributed to optically thin synchrotron emission, suggesting the presence of relativistic electrons.

The potential for very high energy (VHE) gamma emission from extraordinary giant magnetar flares has been discussed, for example, in the context of SGR\,1806$-$20 as a possible hadronic accelerator~\citep{Ioka:2005}. In the case of accelerated electrons, the strong magnetic field of the emission region could induce electron$-$positron pair cascades that can quench any VHE gamma-ray emission. This internal absorption of VHE photons can be avoided if the emission region is located far away from the surface of the magnetar. This is realized, for example, in the trapped fireball scenario~\citep{Thompson_2001}. So far, searches for VHE emission (variable or persistent) from the magnetars SGR\,1806$-$20, 4U\,0142+61 and 1E\,2259+5864 have not revealed any detections \citep{MAGIC-magnetars:2013,HESS-SGR1806:2018}. 


\subsection{Fast radio bursts and magnetars}
Fast radio bursts (FRBs) are powerful radio pulses with a duration of several milliseconds with high brightness temperatures suggesting a coherent emission mechanism~\citep{Petroff:2019tty}. It is believed that FRBs are of extragalactic origin. Over the past few years a rapidly growing number of FRBs have been detected in the radio band, including repeating ones~\citep{Petroff:2016tcr}. Various theoretical emission and source models were put forward since the first detection of these enigmatic bursts~\citep{2020Natur.587...45Z}. Magnetars are proposed as sources of FRBs~\citep{Popov:2007uv}. For example, some magnetars could produce FRBs through strongly magnetized pulses that interact with the material in the surrounding nebula and produce synchrotron maser emission~\citep[e.g.][]{Lyubarsky:2014jta,Metzger:2019,Beloborodov:2020}. Some models suggest that repeating FRBs are generated not far from the surface of the magnetar through ultrarelativistic internal shocks and blast waves in the magnetar wind associated with flares~\citep[e.g.][]{2017ApJ...843L..26B}.

\subsection{Past observations of FRBs with Imaging Atmospheric Cerenkov Telescopes}
The High Energy Stereoscopic System (H.E.S.S.) is an array of one 28 m and four 12m imaging atmospheric Cerenkov
telescopes (IACTs) located in the Khomas Highland in Namibia at an altitude of 1835 m. It is capable of detecting VHE gamma rays from energies of a few tens of GeV to 100 TeV. In the past, H.E.S.S. targeted two FRBs, FRB\,20150215~\citep{2017MNRAS.469.4465P} and FRB\,20150418~\citep{2017A&A...597A.115H} with several hours to days of delay. No significant VHE emission was found from either observations. Simultaneous observations of repeating and nonrepeating FRBs with IACTs and radio telescopes did not reveal gamma-ray counterparts: VERITAS observed FRB\,20121102 and FRB\,180814.J0422+73 simultaneously with the Green Bank Telescope, detecting 15 radio bursts during the campaign~\citep{2019ICRC...36..698H}. MAGIC observed FRB\,20121102 simultaneously with Arecibo, which detected five radio bursts~\citep{Acciari:2018hnf}. 


\subsection{SGR~1935+2154}
Following its discovery in 2014~\citep{Swift_discovery}, SGR\,1935+2154 has probably become the most burst-active SGR, emitting dozens of X-ray bursts over the past few years~\citep{Lin:2020-I}. SGR\,1935+2154 is associated with the middle-aged galactic SNR\,G57.2+0.8 at a distance of about 6.6~kpc~\citep{Zhou:2020}. In 2020 late April and May SGR\,1935+2154 showed renewed X-ray burst activity culminating with a ``forest of bursts" as detected by BAT on board the Neil Gehrels Swift Observatory. Many other X-ray and soft gamma-ray telescopes (Fermi-GBM, INTEGRAL, sAGILE, HXMT, Konus–Wind, NICER) also reported sustained bursting activity into late May.

This situation became considerably more interesting with the detection of short, intense radio bursts from the direction of SGR\,1935+2154. Two millisecond-duration radio bursts (FRB\,20200428) were detected, the first burst being a double-peaked one detected by CHIME and STARE2~\citep{SGR1935_STAR2, SGR1935_CHIME} on 2020 April 28, at 14:34:33 UTC and the second burst by FAST \citep{FAST_FRB} on April 30, at 21:43:00 UTC. The radio energy released under the assumption of an isotropic emission of the bursts at 6.6 kpc is about $10^{34}$-$10^{35}$\,erg, just below the low end of the extragalactic FRB distribution observed so far. After removing the radio dispersion delay, the timing of the first radio burst appears to line up very well with one of the bright X-ray bursts seen by AGILE~\citep{tavani2020xray}, Konus-Wind~\citep{Ridnaia:2021}, INTEGRAL~\citep[burst G in][]{INTEGRAL_BURST_A} and Insight-HXMT~\citep{HXMT_Bursts,HXMT_Delay}. This coincidence is the first evidence that magnetars are linked to FRBs, or at least a subset of (repeating) bursts. It is also shown that the X-ray bursts overlapping the double-peaked CHIME radio burst have an unusually hard spectrum, and it is suggested that these X-rays and the radio bursts arise from a common scenario \citep{Ridnaia:2021}. Furthermore, the nonthermal nature of the Insight-HXMT burst~\citep{li2020insighthxmt} points to the production of multi-TeV electrons. Multi-GeV to TeV gamma-ray emission via the inverse-Compton process may then accompany this X-ray emission.


We here report on searches for VHE gamma-ray emission associated with the flares of SGR\,1935+2154 with H.E.S.S. during the period of high activity on 2020 April 28. 
This paper is organized as follows: Sec.~\ref{sec:SGR1935_observations} describes the H.E.S.S. observations and provides an overview of MWL observations of SGR\,1935+2154. Sec.~\ref{sec:SGR1935_hess_analysis} presents the H.E.S.S. data analysis and results. Sec.~\ref{sec:SGR1935_discussion} discusses our findings, which are concluded in Sec.~\ref{sec:SGR1935_conclusion}.

\section{Observation summary}
\label{sec:SGR1935_observations}

\begin{figure*}[!tb]
\centering
\includegraphics[width=1.0\linewidth]{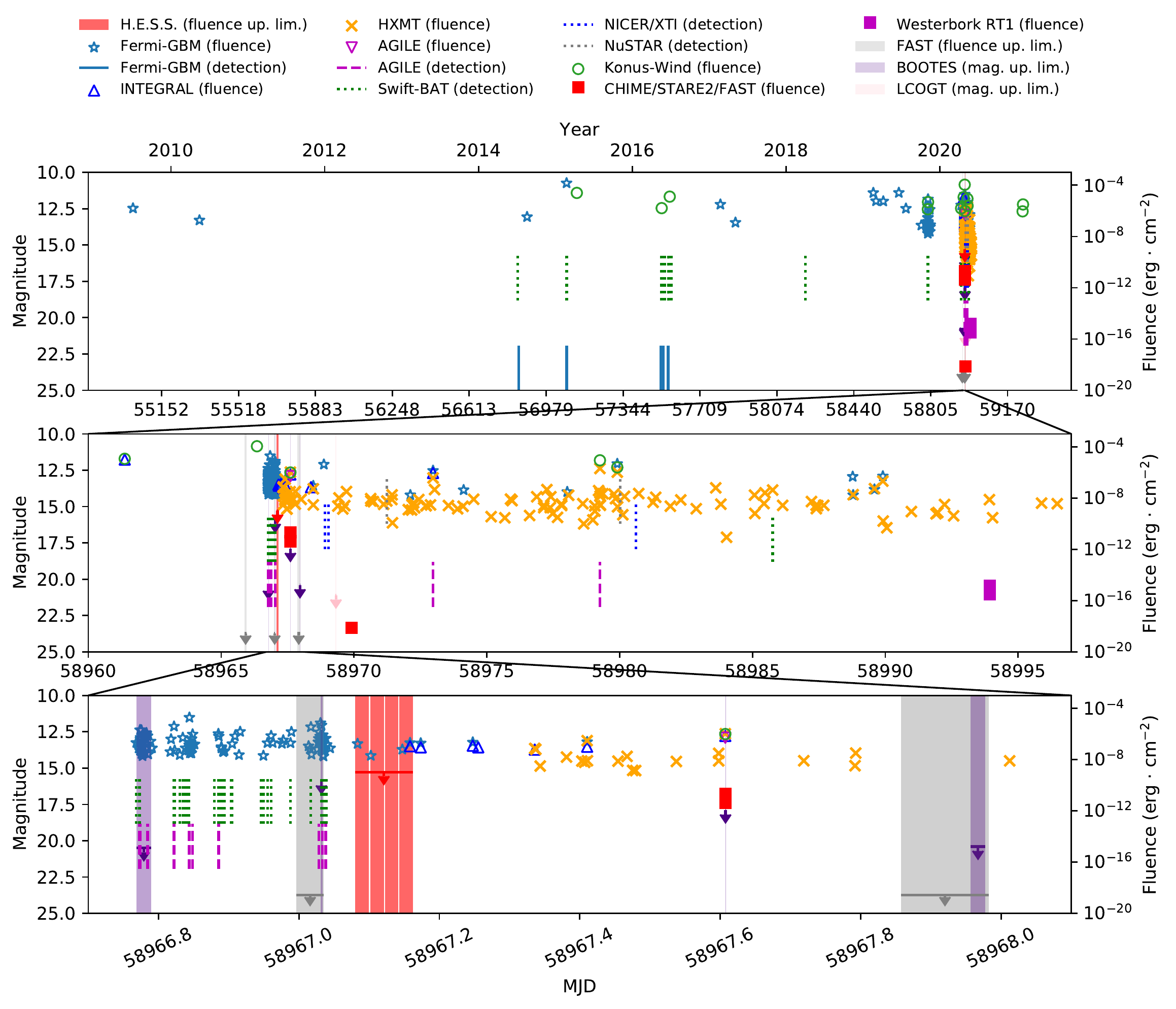}
\caption{SGR\,1935+2154 observations with gamma-ray, X-ray optical and radio telescopes.This plot presents X-ray bursts from the source detected by Fermi-GBM~\citep{Lin_2020,SGR1935_GBM1,SGR1935_GBM2,SGR1935_GBM3,SGR1935_GBM4,SGR1935_GBM5,SGR1935_GBM6,von_Kienlin_2020,Gruber_2014,von_Kienlin_2014,Bhat_2016}, Swift-BAT~\citep{Swift-BAT_ATel}, NICER/XTI, NuSTAR~\citep{Borghese_2020}, INTEGRAL~\citep{INTEGRAL_BURST_A}, HXMT~\citep{HXMT_Bursts2,HXMT_Bursts}, AGILE~\citep{AGILE_ATEL,tavani2020xray} and Konus-Wind~\citep{Ridnaia:2021,SGR1935_KONUS-WIND2,SGR1935_KONUS-WIND3,SGR1935_KONUS-WIND4,SGR1935_KONUS-WIND6,SGR1935_KONUS-WIND7,SGR1935_KONUS-WIND8,SGR1935_KONUS-WIND9,SGR1935_KONUS-WIND10,SGR1935_KONUS-WIND11,SGR1935_KONUS-WIND12,SGR1935_KONUS-WIND13}; radio burst from CHIME~\citep{SGR1935_CHIME,SGR1935_CHIME2}, STARE2~\citep{SGR1935_STAR2}, FAST~\citep{FAST_FRB} and the Westerbork (RT1), Onsala (25m), Toruń (30m) dishes~\citep{Kirsten_2020}. Plot also shows the H.E.S.S., FAST, BOOTES  and LCOGT observations~\citep{2020Natur.587...63L} with upper limits from all shown instruments.}
\label{fig:SGR1935_obs}
\end{figure*}

The top  panel of Fig.~\ref{fig:SGR1935_obs}, gives an overview of the MWL bursts detected from SGR\,1935+2154 over the years. The middle panel zooms in on the 2020 active period. The bottom panel zooms in on the period around the H.E.S.S. observations. The observations during the active period are summarized in the following.

Gamma rays and X-rays: Fig.~\ref{fig:SGR1935_obs} (and references therein) shows detected bursts with AGILE, Fermi-GBM, INTEGRAL, Konus-Wind, HXMT, NuSTAR, NICER/XTI and Swift-BAT from 2009 until the end of 2020. From the Konus-Wind detected burst clusters on 2020 April 27~\citep{SGR1935_KONUS-WIND9} we show only the most intense bursting activity that occurred around 18:33:01 (58965.77293 MJD) with a duration of $\sim$23\,s and a very high fluence of $1.09 \times 10^{-4}\, \mathrm{erg}\,\mathrm{cm}^{-2}$. 

Four Fermi-GBM bursts occurred during the H.E.S.S. follow-up observations of the source~\citep{Lin_2020}. Moreover, INTEGRAL reported the detection of an X-ray burst at 03:47:52.2 UTC (burst A in~\citet{INTEGRAL_BURST_A} at 58967.15824306\,MJD) that coincides with the fourth Fermi-GBM burst during the last data-taking run by H.E.S.S. Therefore, the H.E.S.S. observations provide for the first time simultaneous VHE gamma-ray observational data with X-ray bursts emanating from a SGR.

An X-ray burst is detected by AGILE, Konus-Wind, HXMT and INTEGRAL~\citep[burst G in][]{INTEGRAL_BURST_A} simultaneously to the FRB from CHIME and STARE2 (FRB\,20200428).

Radio: FRB\,20200428 associated with SGR\,1935+2154 is detected by CHIME~\citep{SGR1935_CHIME,SGR1935_CHIME2} at 2020 April 28 14:34:33 UTC and STARE2~\citep{SGR1935_STAR2} at 2020 April 28 14:34:25 UTC ($\sim$58967.60733\,MJD). Several X-ray instruments detected coincident X-ray bursts as shown in Fig.~\ref{fig:SGR1935_obs}. Follow-up observations by the FAST radio telescope (Pingtang, China) in the 1.25\,GHz band did not reveal radio bursts down to a fluence of $<22\,\mathrm{mJy\,\,ms}$~\citep{2020Natur.587...63L}; however, a weak, highly linearly polarized radio burst was detected on April~30~\citep{FAST_FRB} (58969\,MJD). Moreover, no radio bursts were detected in the observation campaigns by the  Arecibo, Effelsberg, LOFAR, MeerKAT, MK2 and Northern Cross radio telescopes (not shown in Fig.~\ref{fig:SGR1935_obs}) reported in~\citet{Bailes_2021}. The upper limits derived from these observations are between 18\,mJy and 25\,mJy. Two additional radio bursts separated by $\mathrm{~\sim 1.4\,s}$ on 2020 May 24 (\citet{Kirsten_2020};58993\,MJD) were detected following an X-ray burst detected by HXMT during a joint campaign between the Westerbork (RT1), Onsala (25 m), and Toruń (30 m) radio dishes. The burst fluences are several orders of magnitude lower than the two bursts detected by CHIME and STARE2 with values of $\mathrm{112 \pm 22\,Jy\,ms}$ and $\mathrm{24\pm 5\,Jy\,ms}$, respectively. The four FRBs detected from SGR\,1935+2154 thus span around seven orders of magnitude in fluence. No FRBs were detected during the time of the H.E.S.S. data acquisition. 

VHE gamma rays: The H.E.S.S. transients' follow-up system triggered Target of Opportunity follow-up observations on SGR\,1935+2154 after the reception of a first Swift-BAT alert, indicating a high-intensity X-ray burst from SGR\,1935+2154 at 2020 Aril 27 18:26:19.95 UTC (58966.76828646\,MJD). A second Swift-BAT alert arrived $\sim$6.5 minutes later~\citep{SWIFT_1_2}. Darkness and visibility constraints only allowed follow-up observations to commence $\sim$7.5 hr later. The observations lasted 2 hr and consisted of four runs taken with {\it wobble} offsets whereby the source is alternately offset by 0.5 deg in opposite directions of the R.A. and decl. The positions reported by Swift-BAT has a 3' uncertainty and are thus fully comprised within the H.E.S.S. field of view of 2.5 deg radius. A summary of the observations is presented in Tab.~\ref{tab:SGR1935_OBS} and we put them in context with the MWL observations of the burst activity of SGR\,1935+2154 in Fig.~\ref{fig:SGR1935_obs}.

\begin{table}[!hb]
    \centering
    \begin{tabular}{ccc}
    \toprule
    Start time (UTC) &  Duration  & Average Zenith Angle \\ 
    \midrule
    2020 April 28 01:55:00  & 28 minutes & 55.0 deg\\
    2020 April 28 02:26:55  & 28 minutes & 51.0 deg  \\
    2020 April 28 02:56:08  & 28 minutes & 48.1 deg \\
    2020 April 28 03:25:24  & 28 minutes & 46.2 deg \\
    \bottomrule
    \end{tabular}
    \caption{Summary of the H.E.S.S. observations of SGR\,1935+2154. \textbf{Note:} The observations overlapped with magnetar bursts detected by INTEGRAL and Fermi-GBM.}
    \label{tab:SGR1935_OBS}
\end{table}

Optical: No optical emission has been detected by the BOOTES-3 (New Zealand) telescopes observing contemporaneously to the first detected FRB, and $3\sigma$ upper limits in the Z-band are given as $>17.9$\,mag during this epoch~\citep{2020Natur.587...63L}. No optical emission was seen by LCOGT (California, USA), other BOOTES telescopes or the MeerLICHT (not shown in Fig.~\ref{fig:SGR1935_obs}) optical telescope~\citep{Bailes_2021}.

\section{Data Analysis and Results}
\label{sec:SGR1935_hess_analysis}
\begin{figure}
  \centering
\includegraphics[width=1.0\linewidth]{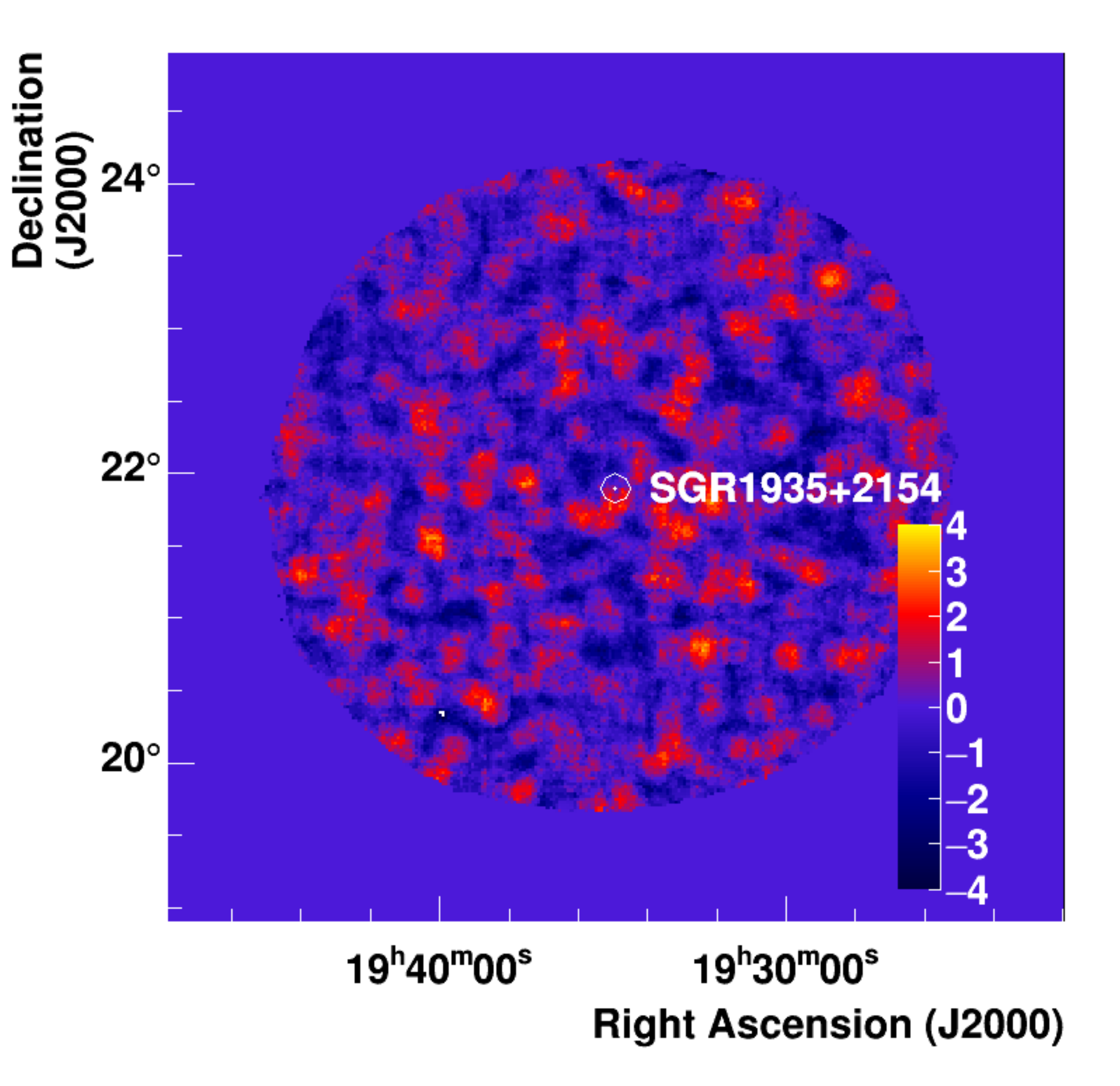}
\caption{Significance map computed from the H.E.S.S. observational data taken on SGR\,1935+2154.}
\label{fig:SGR1935_SIG}
\end{figure}

The H.E.S.S. results presented here use data from all four 12\,m telescopes. The quality of the obtained data has been thoroughly verified and standard quality selection criteria~\citep{Aharonian2006a} are applied. The data are then analyzed using the \textit{standard cuts} of the semi-analytical \textit{Model Analysis} described in~\citet{de-Naurois2009a}. The standard \textit{Ring background} technique~\citep{RingBg} is used to determine the background with a radial acceptance and a $\mathrm{0.1\,deg}$ ON region. We obtain a total number of ON events $\mathrm{N_{ON} = 26}$ and OFF events $\mathrm{N_{OFF} = 270}$ in the source region, with $\mathrm{\alpha = 0.0813}$ leading to an excess value of $4.0$. We derive the significance map shown in Fig.~\ref{fig:SGR1935_SIG} using the formalism described in~\citet{LiMa} and an oversampling radius of $0.1~\mathrm{deg}$. In this map no significant signal above $5\sigma$ can be found at the position of the SGR or elsewhere in the covered region. We therefore conclude that no significant VHE gamma-ray emission has been detected by H.E.S.S. during the follow-up observations of SGR\,1935+2154.

\begin{figure}[htp]
  \centering
\includegraphics[width=1.0\linewidth]{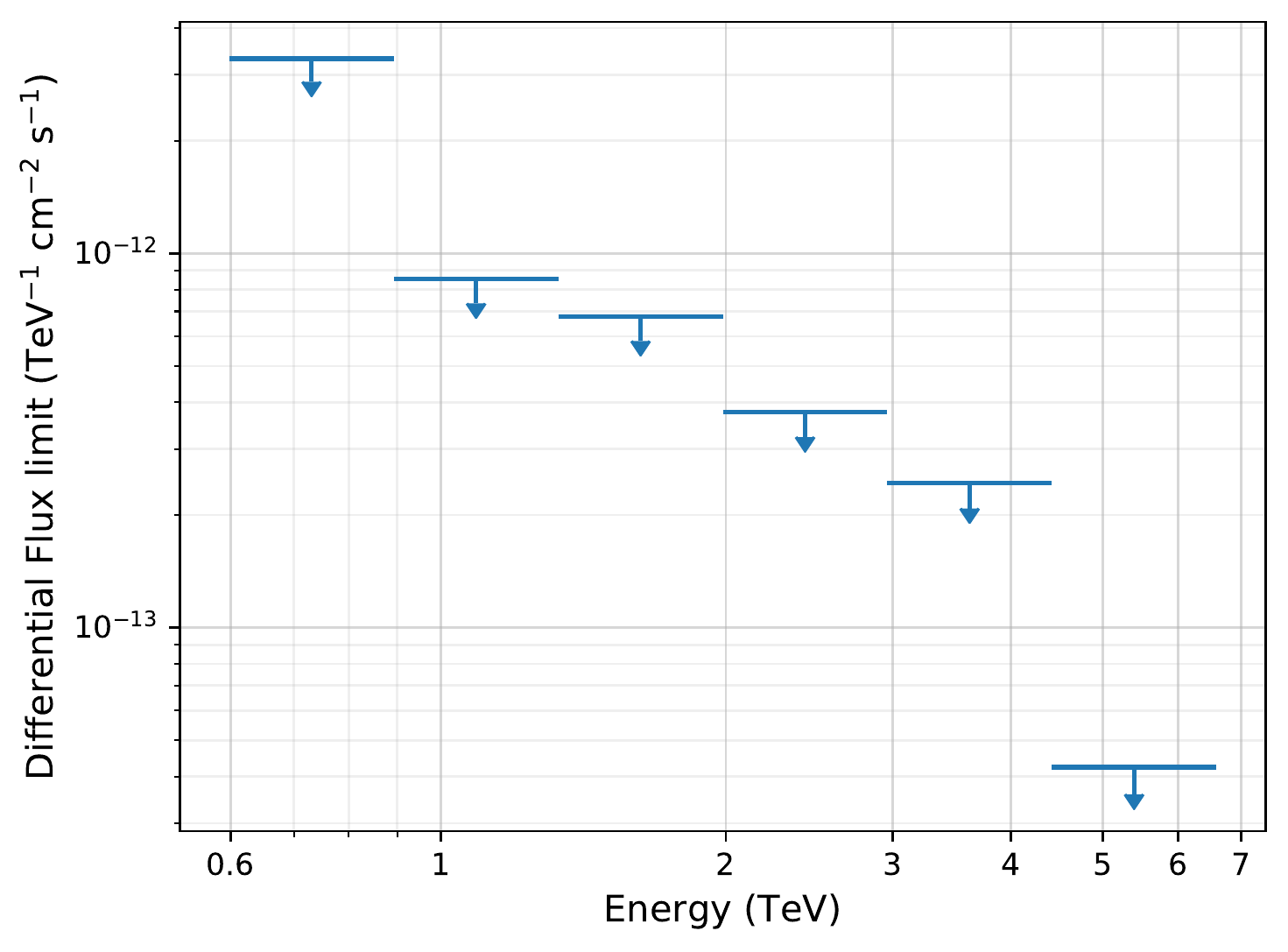}
\caption{Differential 95\% C.L. upper limits derived from the H.E.S.S. observational data taken on SGR\,1935+2154.}
\label{fig:SGR1935_DiffUL}
\end{figure}

A low-energy threshold is defined as the energy where the effective area is at least 10\% of its maximum value. Influenced by the relatively high zenith angle of the observations, a value of $E_{thr}\mathrm{ = 600 \, GeV}$ is found. Assuming a generic $E^{-2.5}$ energy spectrum we compute 95\% confidence level differential upper limits shown in Fig.~\ref{fig:SGR1935_DiffUL} at the position of SGR\,1935+2154 using a Poisson likelihood method described in~\citet{Rolke}. 
Integrating above 600 GeV gives a value of ${\Phi_{\gamma}\mathrm{(E > 600\,GeV)} < 1.5\cdot10^{-12}\,\mathrm{cm}^{-2}\,\mathrm {s}^{-1}}$. If we consider an $E^{-2}$ spectrum, this value drops by $\sim$13\% and for a $E^{-3}$ spectrum it increases by $\sim$7\%. The analysis presented in this section has been cross-checked and validated with an independent event calibration and reconstruction analysis~\citep{Parsons2014a}. 

\begin{figure}
  \centering
\includegraphics[width=1.0\linewidth]{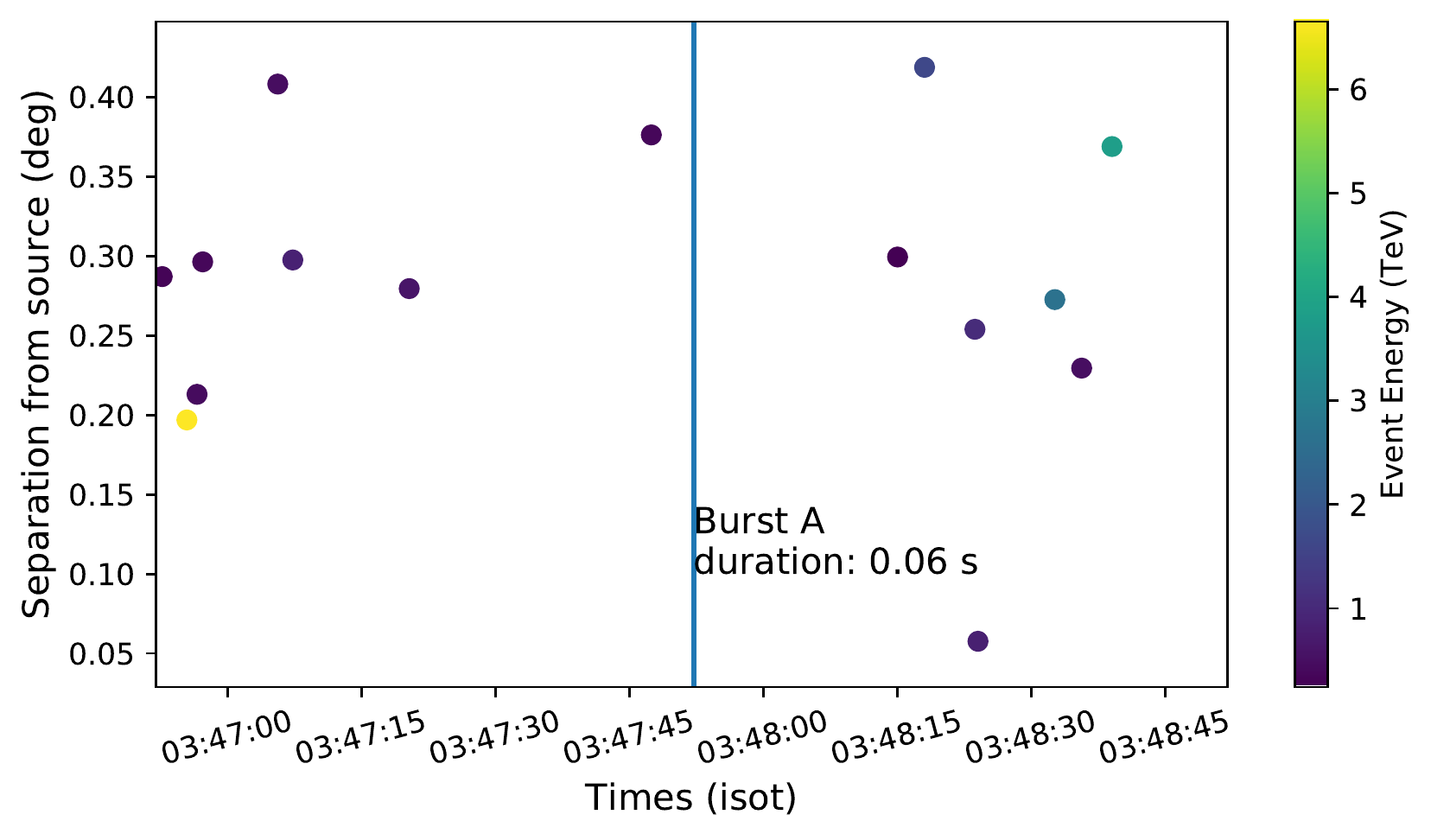}
\caption{VHE gamma candidates detected by H.E.S.S. from SGR1935+2154 at the time of the X-ray burst A detected by INTEGRAL~\citep{INTEGRAL_BURST_A}. The vertical line designates the burst. The gamma candidates are expected background events.}
\label{fig:SGR1935_GammaLike}
\end{figure}

In addition to the standard analysis, we perform the Cumulative Sum, ON-OFF and Exp- tests described in~\citet{Brun_transient_tools} to search for a variable or transient VHE signal. For that we use the gamma-candidate events that are selected after the cut applied by the \textit{Model Analysis}. No significant variability was detected at minute to hour timescales. Furthermore, a search for gamma-candidate doublets arriving within millisecond time windows from the direction of SGR\,1935+2154, which could be associated with a magnetar burst, is conducted. No such doublets are detected. This search is extended to clusters of gamma candidates around the Fermi-GBM and INTEGRAL bursts. As an example, Fig.~\ref{fig:SGR1935_GammaLike} shows the arrival times of gamma candidates around the INTEGRAL burst. No gamma-candidate events from the source region are found within less than 9 s of any Fermi-GBM or INTEGRAL burst occurring during our observations. We note that the gamma candidates are expected background events that passed the selection cuts from the \textit{Model Analysis} and should not be misinterpreted with gamma rays. 

The H.E.S.S. sensitivity for this kind of fast transient phenomena, i.e. assuming detection of gamma-candidate multiplets at millisecond-scale time windows, is at the order of ${\sim 10^{-9}}$\,erg$\,$cm$^{-2}$ depending on the zenith angle of the observations. 

\section{Discussion}
\label{sec:SGR1935_discussion}

The H.E.S.S. observations described here present the first VHE gamma-ray observations of a magnetar during a high-activity phase. Our observations are coincident with four X-ray bursts detected by two instruments (Fermi-GBM and INTEGRAL) and put stringent upper limits on the VHE emission during the active phase of the magnetar.

With a spectral index of 2, the INTEGRAL burst~\citep[burst G from][]{INTEGRAL_BURST_A} coincident with the FRB detected by CHIME and STARE2 has a harder spectral shape than from other bursts detected by INTEGRAL. It is also more energetic than burst A that occurred during the H.E.S.S. observations: ${F_G/F_A = 5.3}$ with $F_G$ and $F_A$ the fluence values of bursts A and G, respectively. Assuming that burst A could be connected to another coincidental FRB half an order of magnitude less energetic than the detected radio burst, FRB\,20200428, this hypothetical FRB would still be within the sensitivity of radio instruments. The fact that several X-ray bursts happened during FAST radio observations and the lack of radio burst detection (${\mathrm{UL} \sim 2.75 \times 10^{-19}\, \mathrm{erg}\,\,\mathrm{cm}^{-2}}$) leads us to believe that these ordinary SGR bursts (including burst A) are different from burst G. Moreover, the HXMT burst associated with the FRB is dominated by a power law and is therefore primarily nonthermal in nature according to \citet{li2020insighthxmt}, which is very rare for SGRs (6\% of SGR bursts as per~\citet{li2020insighthxmt}).

The lack of VHE gamma rays is consistent with expectations from regular magnetar bursts~\citep{Kaspi:2017fwg} when thermal emission mechanism is involved. The integral upper limits derived from the 2 hr of H.E.S.S. observation, assuming a spectral index of $E^{-2.5}$, can be translated into upper limits on the flux $F\mathrm{(E > 600\,GeV)} < 2.4 \times 10^{-12} \,\mathrm{erg} \, \mathrm{cm}^{-2}\,\mathrm{s}^{-1}$. Assuming a distance of 6.6 kpc we derive a luminosity upper limit  $L\mathrm{(E > 600\,GeV)} < 1.3 \times 10^{34} \,\mathrm{erg}\,\mathrm{s}^{-1}$. This places constraints on persistent VHE emission from SGR\,1935+2154 during the H.E.S.S. observations. The sensitivity to gamma-ray multiplets can be transformed into sensitivity to the isotropic energy from a VHE burst $\mathrm{E_{VHE,iso}\mathrm{(E > 600\,GeV)} \leq 5.2 \times 10^{36} \,erg}$. This sensitivity is higher than the isotropic energies of the FRB bursts detected from the source ($\mathrm{\sim10^{34} \,erg - 10^{35}\, erg}$). However, it can be compared to the energy released in the X-ray domain during the coincident Fermi-GBM and INTEGRAL bursts ($\mathrm{\sim10^{38} \,erg - 10^{39}\, erg}$), indicating that if there were an isotropic VHE emission related to these X-ray bursts  during the time of H.E.S.S. observations it would have been detected.

The nondetection by H.E.S.S. may suggest that the inverse-Compton process is suppressed in the magnetar surroundings, making the VHE emission too weak to be detected. An explanation for that is that the gamma-ray emission is happening too close to the magnetar surface and pair production and photon splitting result in significant energy losses for the VHE gamma rays, leading to strong cutoffs in the MeV to GeV energy range. This flux suppression could be avoided in scenarios where the gamma rays are generated well away from the magnetar's intense magnetic field~\citep{Hu_2019} or in scenarios involving axions~\citep{Archer:2020znv}. In case of detection, the H.E.S.S. observations could therefore also probe the particle transport aspects (such as outflows) in the vicinity of SGR\,1935+2154 during the recent flaring episode. Further MWL observations of flaring magnetars are needed for a robust conclusion on the nature of the observed bursts. We cannot draw any conclusions concerning VHE counterpart emissions from FRBs since the H.E.S.S. data presented here are not contemporaneous with any radio burst and the coincident X-ray bursts seem to be different in nature from the bursts associated with FRB\,20200428. The predicted energy released during a VHE pulse from magnetar nebulae by~\citet{Lyubarsky:2014jta} is detectable at distances of roughly hundreds Mpc. 
The derived sensitivity of H.E.S.S. to TeV photons suggests that such VHE bursts would be detectable by H.E.S.S. This motivates further observations of magnetars and FRBs in the VHE domain. 


\section{Conclusions}
\label{sec:SGR1935_conclusion}
H.E.S.S. observed SGR\,1935+2154 during its period of high activity on 2020 April 28. We gather MWL information on the source from the time of discovery until the high-activity period recorded in 2020. Four X-ray bursts were detected by INTEGRAL and Fermi-GBM during the H.E.S.S observation period. The data analysis does not show any significant detection of VHE gamma rays from the source and variability searches do not show emission of gamma rays on the minutes, seconds and milliseconds scales. We thus use H.E.S.S. observations to derive upper limits. SGR\,1935+2154 established the link between FRBs and X-ray bursts thanks to the coincident radio and X-ray bursts. While the H.E.S.S. upper limits cannot be used to constrain VHE emission from the source's FRBs, they place for the first time constraining upper limits on the persistent and transient VHE emission coincident with magnetar X-ray bursts. The details of the underlying emission mechanism are still unclear and further MWL observations of these objects are necessary. We note that SGR\,1935+2154 renewed its activity in the beginning of 2021 with bursts detected by INGTERAL~\citep{INTEGRAL_2021}, Fermi-GBM~\citep{SGR1935_GBM_2021-1}~\citep{SGR1935_GBM_2021-2}, CGBM~\citep{CGBM_2021} and Konus-Wind~\citep{SGR1935_KONUS-WIND_2021}. The H.E.S.S. program on transient phenomena includes various triggered and untriggered campaigns and aims at providing additional pieces to the puzzle of the origin of FRBs and the mechanisms of magnetar flares.

\section*{acknowledgments}
The support of the Namibian authorities and of the University of Namibia in facilitating the construction and operation of H.E.S.S. is gratefully acknowledged, as is the support by the German Ministry for Education and Research (BMBF), the Max Planck Society, the German Research Foundation (DFG), the Helmholtz Association,  the Alexander von Humboldt Foundation, the French Ministry of Higher  Education,  Research and Innovation, the Centre National de la Recherche Scientifique (CNRS/IN2P3 and CNRS/INSU), the Commissariat a l'\'energie atomique et aux \'energies alternatives (CEA), the U.K. Science and Technology Facilities Council (STFC), the Knut and Alice Wallenberg Foundation, the National Science Centre,  Poland grant No. 2016/22/M/ST9/00382, the South African Department of Science and Technology and National Research Foundation, the University of Namibia, the National Commission on Research, Science \& Technology of Namibia (NCRST), the Austrian Federal Ministry of Education, Science and Research and the Austrian Science Fund (FWF), the Australian Research Council (ARC), the Japan Society for the Promotion of Science and by the University of Amsterdam. We appreciate the excellent work of the technical support staff in Berlin, Zeuthen, Heidelberg, Palaiseau, Paris, Saclay, T\"ubingen and in Namibia in the construction and operation of the equipment. This work benefited from services provided by the H.E.S.S. Virtual Organisation, supported by the national resource providers of the EGI Federation.
We thank Sergei Popov, Konstantin Postnov and Maxim Pshirkov for fruitful discussions on the interpretation of the observations. 

\bibliography{sample631}{}
\bibliographystyle{aasjournal}

\allauthors

\end{document}

%% file: authors_apj.tex
\collaboration{0}{H.E.S.S. Collaboration}

\author{H.~Abdalla}
\affiliation{University of Namibia, Department of Physics, Private Bag 13301, Windhoek 10005, Namibia}
\author[0000-0003-1157-3915]{F.~Aharonian}
\affiliation{Dublin Institute for Advanced Studies, 31 Fitzwilliam Place, Dublin 2, Ireland,Max-Planck-Institut f\"ur Kernphysik, P.O. Box 103980, D 69029 Heidelberg, Germany}
\affiliation{High Energy Astrophysics Laboratory, RAU,  123 Hovsep Emin St  Yerevan 0051, Armenia}
\author{F.~Ait~Benkhali}
\affiliation{Max-Planck-Institut f\"ur Kernphysik, P.O. Box 103980, D 69029 Heidelberg, Germany}
\author{E.O.~Ang\"uner}
\affiliation{Aix Marseille Universit\'e, CNRS/IN2P3, CPPM, Marseille, France}
\author[0000-0002-1998-9707]{C.~Arcaro}
\affiliation{Centre for Space Research, North-West University, Potchefstroom 2520, South Africa}
\author{C.~Armand}
\affiliation{Laboratoire d'Annecy de Physique des Particules, Univ. Grenoble Alpes, Univ. Savoie Mont Blanc, CNRS, LAPP, 74000 Annecy, France}
\author[0000-0001-5067-2620]{T.~Armstrong}
\affiliation{University of Oxford, Department of Physics, Denys Wilkinson Building, Keble Road, Oxford OX1 3RH, UK}
\author[0000-0002-2153-1818]{H.~Ashkar}
\affiliation{IRFU, CEA, Universit\'e Paris-Saclay, F-91191 Gif-sur-Yvette, France}
\author[0000-0002-9326-6400]{M.~Backes}
\affiliation{University of Namibia, Department of Physics, Private Bag 13301, Windhoek 10005, Namibia}
\affiliation{Centre for Space Research, North-West University, Potchefstroom 2520, South Africa}
\author[0000-0003-0477-1614]{V.~Baghmanyan}
\affiliation{Instytut Fizyki J\c{a}drowej PAN, ul. Radzikowskiego 152, 31-342 Krak{\'o}w, Poland}
% ...


%% file: main.bbl
\begin{thebibliography}{}
\expandafter\ifx\csname natexlab\endcsname\relax\def\natexlab#1{#1}\fi
\providecommand{\url}[1]{\href{#1}{#1}}
\providecommand{\dodoi}[1]{doi:~\href{http://doi.org/#1}{\nolinkurl{#1}}}
\providecommand{\doeprint}[1]{\href{http://ascl.net/#1}{\nolinkurl{http://ascl.net/#1}}}
\providecommand{\doarXiv}[1]{\href{https://arxiv.org/abs/#1}{\nolinkurl{https://arxiv.org/abs/#1}}}

\bibitem[{{Acciari} {et~al.}(2018){Acciari}, {Ansoldi}, {Antonelli}, {Arbet
  Engels}, {Arcaro}, {Baack}, {Babi{\'c}}, {}, {Banerjee}, {Bangale}, {Barres
  de Almeida}, {Barrio}, {Becerra Gonz{\'a}lez}, {Bednarek}, {Bernardini},
  {Berti}, {Besenrieder}, {Bhattacharyya}, {Bigongiari}, {Biland}, {Blanch},
  {Bonnoli}, {Carosi}, {Ceribella}, {Chatterjee}, {Colak}, {Colin}, {Colombo},
  {Contreras}, {Cortina}, {Covino}, {Cumani}, {D'Elia}, {da Vela}, {Dazzi}, {de
  Angelis}, {de Lotto}, {Delfino}, {Delgado}, {di Pierro}, {Dom{\'\i}nguez},
  {Dominis Prester}, {Dorner}, {Doro}, {Einecke}, {Elsaesser}, {Fallah
  Ramazani}, {Fattorini}, {Fern{\'a}ndez-Barral}, {Ferrara}, {Fidalgo},
  {Foffano}, {Fonseca}, {Font}, {Fruck}, {Gallozzi}, {Garc{\'\i}a L{\'o}pez},
  {Garczarczyk}, {Gaug}, {Giammaria}, {Godinovi{\'c}}, {}, {Guberman},
  {Hadasch}, {Hahn}, {Hassan}, {Herrera}, {Hoang}, {Hrupec}, {Inoue}, {Ishio},
  {Iwamura}, {Kubo}, {Kushida}, {Kuve{\v{z}}di{\'c}}, {}, {Lamastra}, {Lelas},
  {Leone}, {Lindfors}, {Lombardi}, {Longo}, {L{\'o}pez}, {L{\'o}pez-Oramas},
  {Maggio}, {Majumdar}, {Makariev}, {Maneva}, {Manganaro}, {Mannheim},
  {Maraschi}, {Mariotti}, {Mart{\'\i}nez}, {Masuda}, {Mazin}, {Minev},
  {Miranda}, {Mirzoyan}, {Molina}, {Moralejo}, {Moreno}, {Moretti},
  {Neustroev}, {Niedzwiecki}, {Nievas Rosillo}, {Nigro}, {Nilsson}, {Ninci},
  {Nishijima}, {Noda}, {Nogu{\'e}s}, {Paiano}, {Palacio}, {Paneque},
  {Paoletti}, {Paredes}, {Pedaletti}, {Pe{\~n}il}, {Peresano}, {Persic}, {Prada
  Moroni}, {Prandini}, {Puljak}, {Garcia}, {Rhode}, {Rib{\'o}}, {Rico},
  {Righi}, {Rugliancich}, {Saha}, {Saito}, {Satalecka}, {Schweizer}, {Sitarek},
  {{\v{S}}nidari{\'c}}, {}, {Sobczynska}, {Somero}, {Stamerra}, {Strzys},
  {Suri{\'c}}, {}, {Tavecchio}, {Temnikov}, {Terzi{\'c}}, {}, {Teshima},
  {Torres-Alb{\`a}}, {Tsujimoto}, {Vanzo}, {Vazquez Acosta}, {Vovk}, {Ward},
  {Will}, {Zari{\'c}}, {Marcote}, {Spitler}, {Hessels}, {Kashiyama}, {Murase},
  {Bosch-Ramon}, {Michilli}, \& {Seymour}}]{Acciari:2018hnf}
{Acciari}, V.~A., {Ansoldi}, S., {Antonelli}, L.~A., {et~al.} 2018, \mnras,
  481, 2479, \dodoi{10.1093/mnras/sty2422}

\bibitem[{{Aharonian} {et~al.}(2006){Aharonian}, {Akhperjanian}, {Bazer-Bachi},
  {Beilicke}, {Benbow}, {Berge}, {Bernl{\"o}hr}, {Boisson}, {Bolz}, {Borrel},
  {Braun}, {Breitling}, {Brown}, {B{\"u}hler}, {B{\"u}sching}, {Carrigan},
  {Chadwick}, {Chounet}, {Cornils}, {Costamante}, {Degrange}, {Dickinson},
  {Djannati-Ata{\"\i}}, {O'C.~Drury}, {Dubus}, {Egberts}, {Emmanoulopoulos},
  {Espigat}, {Feinstein}, {Ferrero}, {Fiasson}, {Fontaine}, {Funk}, {Funk},
  {Gallant}, {Giebels}, {Glicenstein}, {Goret}, {Hadjichristidis}, {Hauser},
  {Hauser}, {Heinzelmann}, {Henri}, {Hermann}, {Hinton}, {Hofmann}, {Holleran},
  {Horns}, {Jacholkowska}, {de Jager}, {Kh{\'e}lifi}, {Komin}, {Konopelko},
  {Kosack}, {Latham}, {Le Gallou}, {Lemi{\`e}re}, {Lemoine-Goumard}, {Lohse},
  {Martin}, {Martineau-Huynh}, {Marcowith}, {Masterson}, {McComb}, {de
  Naurois}, {Nedbal}, {Nolan}, {Noutsos}, {Orford}, {Osborne}, {Ouchrif},
  {Panter}, {Pelletier}, {Pita}, {P{\"u}hlhofer}, {Punch}, {Raubenheimer},
  {Raue}, {Rayner}, {Reimer}, {Reimer}, {Ripken}, {Rob}, {Rolland}, {Rowell},
  {Sahakian}, {Saug{\'e}}, {Schlenker}, {Schlickeiser}, {Schwanke}, {Sol},
  {Spangler}, {Spanier}, {Steenkamp}, {Stegmann}, {Superina}, {Tavernet},
  {Terrier}, {Th{\'e}oret}, {Tluczykont}, {van Eldik}, {Vasileiadis}, {Venter},
  {Vincent}, {V{\"o}lk}, {Wagner}, \& {Ward}}]{Aharonian2006a}
{Aharonian}, F., {Akhperjanian}, A.~G., {Bazer-Bachi}, A.~R., {et~al.} 2006,
  A\&A, 457, 899, \dodoi{10.1051/0004-6361:20065351}

\bibitem[{{Aleksić, J.} {et~al.}(2013){Aleksić, J.}, {Antonelli, L. A.},
  {Antoranz, P.}, {Asensio, M.}, {Barres de Almeida, U.}, {Barrio, J. A.},
  {Becerra Gonz\'alez, J.}, {Bednarek, W.}, {Berger, K.}, {Bernardini, E.},
  {Biland, A.}, {Blanch, O.}, {Bock, R. K.}, {Boller, A.}, {Bonnoli, G.},
  {Borla Tridon, D.}, {Bretz, T.}, {Carmona, E.}, {Carosi, A.}, {Colin, P.},
  {Colombo, E.}, {Contreras, J. L.}, {Cortina, J.}, {Cossio, L.}, {Covino, S.},
  {Da Vela, P.}, {Dazzi, F.}, {De Angelis, A.}, {De Caneva, G.}, {De Cea del
  Pozo, E.}, {De Lotto, B.}, {Delgado Mendez, C.}, {Diago Ortega, A.}, {Doert,
  M.}, {Dominis Prester, D.}, {Dorner, D.}, {Doro, M.}, {Eisenacher, D.},
  {Elsaesser, D.}, {Ferenc, D.}, {Fonseca, M. V.}, {Font, L.}, {Fruck, C.},
  {Garc\'{\i}a L\'opez, R. J.}, {Garczarczyk, M.}, {Garrido Terrats, D.},
  {Gaug, M.}, {Giavitto, G.}, {Godinovi\'{}c, N.}, {Gonz\'alez Mu\~noz, A.},
  {Gozzini, S. R.}, {Hadamek, A.}, {Hadasch, D.}, {H\"afner, D.}, {Herrero,
  A.}, {Hose, J.}, {Hrupec, D.}, {Huber, B.}, {Jankowski, F.}, {Jogler, T.},
  {Kadenius, V.}, {Klepser, S.}, {Knoetig, M. L.}, {Kr\"ahenb\"uhl, T.},
  {Krause, J.}, {Kushida, J.}, {La Barbera, A.}, {Lelas, D.}, {Leonardo, E.},
  {Lewandowska, N.}, {Lindfors, E.}, {Lombardi, S.}, {L\'opez, M.},
  {L\'opez-Coto, R.}, {L\'opez-Oramas, A.}, {Lorenz, E.}, {Makariev, M.},
  {Maneva, G.}, {Mankuzhiyil, N.}, {Mannheim, K.}, {Maraschi, L.}, {Marcote,
  B.}, {Mariotti, M.}, {Mart\'{\i}nez, M.}, {Mazin, D.}, {Meucci, M.},
  {Miranda, J. M.}, {Mirzoyan, R.}, {Mold\'on, J.}, {Moralejo, A.},
  {Munar-Adrover, P.}, {Niedzwiecki, A.}, {Nieto, D.}, {Nilsson, K.}, {Nowak,
  N.}, {Orito, R.}, {Paiano, S.}, {Palatiello, M.}, {Paneque, D.}, {Paoletti,
  R.}, {Paredes, J. M.}, {Partini, S.}, {Persic, M.}, {Pilia, M.}, {Pochon,
  J.}, {Prada, F.}, {Prada Moroni, P. G.}, {Prandini, E.}, {Puljak, I.},
  {Reichardt, I.}, {Reinthal, R.}, {Rhode, W.}, {Rib\'o, M.}, {Rico, J.},
  {R\"ugamer, S.}, {Saggion, A.}, {Saito, K.}, {Saito, T. Y.}, {Salvati, M.},
  {Satalecka, K.}, {Scalzotto, V.}, {Scapin, V.}, {Schultz, C.}, {Schweizer,
  T.}, {Shore, S. N.}, {Sillanp\"a\"a, A.}, {Sitarek, J.}, {Snidaric, I.},
  {Sobczynska, D.}, {Spanier, F.}, {Spiro, S.}, {Stamatescu, V.}, {Stamerra,
  A.}, {Steinke, B.}, {Storz, J.}, {Sun, S.}, {Suri\'{}c, T.}, {Takalo, L.},
  {Takami, H.}, {Tavecchio, F.}, {Temnikov, P.}, {Terzi\'{}c, T.}, {Tescaro,
  D.}, {Teshima, M.}, {Tibolla, O.}, {Torres, D. F.}, {Toyama, T.}, {Treves,
  A.}, {Uellenbeck, M.}, {Vogler, P.}, {Wagner, R. M.}, {Weitzel, Q.},
  {Zabalza, V.}, {Zandanel, F.}, {Zanin, R.}, {Rea, N.}, \& {Backes,
  M.}}]{MAGIC-magnetars:2013}
{Aleksić, J.}, {Antonelli, L. A.}, {Antoranz, P.}, {et~al.} 2013, A\&A, 549,
  A23, \dodoi{10.1051/0004-6361/201220275}

\bibitem[{Archer \& Buckley(2021)}]{Archer:2020znv}
Archer, A., \& Buckley, J. 2021, PoS, ICRC2019, 504,
  \dodoi{10.22323/1.358.0504}

\bibitem[{Bailes {et~al.}(2021)Bailes, Bassa, Bernardi, Buchner, Burgay, Caleb,
  Cooper, Desvignes, Groot, Heywood, Jankowski, Karuppusamy, Kramer, Malenta,
  Naldi, Pilia, Pupillo, Rajwade, Spitler, Surnis, Stappers, Addis, Bloemen,
  Bezuidenhout, Bianchi, Champion, Chen, Driessen, Geyer, Gourdji, Hessels,
  Kondratiev, Klein-Wolt, Körding, Le Poole, Liu, Lower, Lyne, Magro,
  McBride, Mickaliger, Morello, Parthasarathy, Paterson, Perera, Pieterse,
  Pleunis, Possenti, Rowlinson, Serylak, Setti, Tavani, Wijers, ter Veen,
  Venkatraman Krishnan, Vreeswijk, \& Woudt}]{Bailes_2021}
Bailes, M., Bassa, C.~G., Bernardi, G., {et~al.} 2021, MNRAS, 503, 5367,
  \dodoi{10.1093/mnras/stab749}

\bibitem[{Barthelmy {et~al.}(2020)Barthelmy, Bernardini, D'Avanzo, Gropp,
  Kennea, Lien, Melandri, Palmer, Sbarratoand, \& Siegel}]{SWIFT_1_2}
Barthelmy, S.~D., Bernardini, M.~G., D'Avanzo, P., {et~al.} 2020, GCN 27657.
\newblock \url{https://gcn.gsfc.nasa.gov/gcn/gcn3/27657.gcn3}

\bibitem[{{Beloborodov}(2017)}]{2017ApJ...843L..26B}
{Beloborodov}, A.~M. 2017, \apjl, 843, L26, \dodoi{10.3847/2041-8213/aa78f3}

\bibitem[{{Beloborodov}(2020)}]{Beloborodov:2020}
---. 2020, \apj, 896, 142, \dodoi{10.3847/1538-4357/ab83eb}

\bibitem[{{Berge} {et~al.}(2007){Berge}, {Funk}, \& {Hinton}}]{RingBg}
{Berge}, D., {Funk}, S., \& {Hinton}, J. 2007, \aap, 466, 1219,
  \dodoi{10.1051/0004-6361:20066674}

\bibitem[{Bhat {et~al.}(2016)Bhat, Meegan, von Kienlin, Paciesas, Briggs,
  Burgess, Burns, Chaplin, Cleveland, Collazzi, Connaughton, Diekmann,
  Fitzpatrick, Gibby, Giles, Goldstein, Greiner, Jenke, Kippen, Kouveliotou,
  Mailyan, McBreen, Pelassa, Preece, Roberts, Sparke, Stanbro, Veres,
  Wilson-Hodge, Xiong, Younes, Yu, \& Zhang}]{Bhat_2016}
Bhat, P.~N., Meegan, C.~A., von Kienlin, A., {et~al.} 2016, \apjs, 223, 28,
  \dodoi{10.3847/0067-0049/223/2/28}

\bibitem[{Bochenek {et~al.}(2020)Bochenek, Kulkarni, Ravi, McKenna, Hallinan,
  \& Belov}]{SGR1935_STAR2}
Bochenek, C., Kulkarni, S., Ravi, V., {et~al.} 2020, ATel, 13684.
\newblock \url{http://www.astronomerstelegram.org/?read=13684}

\bibitem[{Borghese {et~al.}(2020)Borghese, Zelati, Rea, Esposito, Israel,
  Mereghetti, \& Tiengo}]{Borghese_2020}
Borghese, A., Zelati, F.~C., Rea, N., {et~al.} 2020, ApJL, 902, L2,
  \dodoi{10.3847/2041-8213/aba82a}

\bibitem[{Brun {et~al.}(2020)Brun, Piel, de~Naurois, \&
  Bernhard}]{Brun_transient_tools}
Brun, F., Piel, Q., de~Naurois, M., \& Bernhard, S. 2020, APh, 118, 102429,
  \dodoi{10.1016/j.astropartphys.2020.102429}

\bibitem[{Burns \& Younes(2015)}]{SGR1935_GBM2}
Burns, E., \& Younes, G. 2015, GCN 17496.
\newblock \url{https://gcn.gsfc.nasa.gov/gcn3/17496.gcn3}

\bibitem[{{CHIME/FRB Collaboration} {et~al.}(2020){CHIME/FRB Collaboration},
  {Andersen}, {Bandura}, {Bhardwaj}, {Bij}, {Boyce}, {Boyle}, {Brar},
  {Cassanelli}, {Chawla}, {Chen}, {Cliche}, {Cook}, {Cubranic}, {Curtin},
  {Denman}, {Dobbs}, {Dong}, {Fandino}, {Fonseca}, {Gaensler}, {Giri}, {Good},
  {Halpern}, {Hill}, {Hinshaw}, {H{\"o}fer}, {Josephy}, {Kania}, {Kaspi},
  {Landecker}, {Leung}, {Li}, {Lin}, {Masui}, {McKinven}, {Mena-Parra},
  {Merryfield}, {Meyers}, {Michilli}, {Milutinovic}, {Mirhosseini},
  {M{\"u}nchmeyer}, {Naidu}, {Newburgh}, {Ng}, {Patel}, {Pen},
  {Pinsonneault-Marotte}, {Pleunis}, {Quine}, {Rafiei-Ravandi}, {Rahman},
  {Ransom}, {Renard}, {Sanghavi}, {Scholz}, {Shaw}, {Shin}, {Siegel}, {Singh},
  {Smegal}, {Smith}, {Stairs}, {Tan}, {Tendulkar}, {Tretyakov}, {Vanderlinde},
  {Wang}, {Wulf}, \& {Zwaniga}}]{SGR1935_CHIME}
{CHIME/FRB Collaboration}, {Andersen}, B.~C., {Bandura}, K.~M., {et~al.} 2020,
  \nat, 587, 54, \dodoi{10.1038/s41586-020-2863-y}

\bibitem[{Cummings {et~al.}(2014)Cummings, Barthelmy, Chester, \&
  Page}]{Swift_discovery}
Cummings, J.~R., Barthelmy, S.~D., Chester, M.~M., \& Page, K.~L. 2014, ATel,
  6294.
\newblock \url{http://www.astronomerstelegram.org/?read=6294}

\bibitem[{{de Naurois} \& {Rolland}(2009)}]{de-Naurois2009a}
{de Naurois}, M., \& {Rolland}, L. 2009, APh, 32, 231,
  \dodoi{10.1016/j.astropartphys.2009.09.001}

\bibitem[{{Fermi-LAT Collaboration} {et~al.}(2021){Fermi-LAT Collaboration},
  {Atwood}, {Axelsson}, {Baldini}, {Barbiellini}, {Baring}, {Bastieri},
  {Bellazzini}, {Berretta}, {Bissaldi}, {Blandford}, {Bonino}, {Bregeon},
  {Bruel}, {Buehler}, {Burns}, {Buson}, {Cameron}, {Caraveo}, {Cavazzuti},
  {Chen}, {Cheung}, {Chiaro}, {Ciprini}, {Costantin}, {Crnogorcevic}, {Cutini},
  {D'Ammando}, {de la Torre Luque}, {de Palma}, {Digel}, {Di Lalla}, {Di
  Venere}, {Dirirsa}, {Fukazawa}, {Funk}, {Fusco}, {Gargano}, {Giglietto},
  {Gill}, {Giordano}, {Giroletti}, {Granot}, {Green}, {Grenier}, {Griffin},
  {Guiriec}, {Hays}, {Horan}, {J{\'o}hannesson}, {Kerr}, {Kova{\v{c}}evi{\'c}},
  {Kuss}, {Larsson}, {Latronico}, {Li}, {Longo}, {Loparco}, {Lovellette},
  {Lubrano}, {Maldera}, {Manfreda}, {Mart{\'\i}-Devesa}, {Mazziotta},
  {McEnery}, {Mereu}, {Michelson}, {Mizuno}, {Monzani}, {Morselli},
  {Moskalenko}, {Negro}, {Omodei}, {Orienti}, {Orlando}, {Paliya}, {Paneque},
  {Pei}, {Pesce-Rollins}, {Piron}, {Poon}, {Porter}, {Principe}, {Racusin},
  {Rain{\`o}}, {Rando}, {Rani}, {Razzaque}, {Reimer}, {Reimer}, {Parkinson},
  {Scargle}, {Scotton}, {Serini}, {Sgr{\`o}}, {Siskind}, {Spandre}, {Spinelli},
  {Tajima}, {Takahashi}, {Tak}, {Torres}, {Tosti}, {Troja}, {Wadiasingh},
  {Wood}, {Yassine}, {Yusafzai}, \& {Zaharijas}}]{FermiMagnetar:2021}
{Fermi-LAT Collaboration}, Ajello, M., {Atwood}, W.~B., {Axelsson}, M.,
  {et~al.} 2021, NatAs, 5, 385, \dodoi{10.1038/s41550-020-01287-8}

\bibitem[{Frederiks {et~al.}(2016)Frederiks, Golenetskii, Aptekar, Oleynik,
  Ulanov, Svinkin, Tsvetkova, Lysenko, Kozlova, \&
  Cline}]{SGR1935_KONUS-WIND13}
Frederiks, D., Golenetskii, S., Aptekar, R., {et~al.} 2016, GCN 19613.
\newblock \url{https://gcn.gsfc.nasa.gov/gcn3/19613.gcn3}

\bibitem[{Golenetskii {et~al.}(2015)Golenetskii, Aptekar, Frederiks, Pal'shin,
  Oleynik, Ulanov, Svinkin, Tsvetkova, Lysenko, \&
  Cline}]{SGR1935_KONUS-WIND11}
Golenetskii, S., Aptekar, R., Frederiks, D., {et~al.} 2015, GCN 17703.
\newblock \url{https://gcn.gsfc.nasa.gov/gcn3/17703.gcn3}

\bibitem[{Gruber {et~al.}(2014)Gruber, Goldstein, von Ahlefeld, Bhat, Bissaldi,
  Briggs, Byrne, Cleveland, Connaughton, Diehl, Fishman, Fitzpatrick, Foley,
  Gibby, Giles, Greiner, Guiriec, van~der Horst, von Kienlin, Kouveliotou,
  Layden, Lin, Meegan, McGlynn, Paciesas, Pelassa, Preece, Rau, Wilson-Hodge,
  Xiong, Younes, \& Yu}]{Gruber_2014}
Gruber, D., Goldstein, A., von Ahlefeld, V.~W., {et~al.} 2014, \apjs, 211, 12,
  \dodoi{10.1088/0067-0049/211/1/12}

\bibitem[{{H.~E.~S.~S. Collaboration} {et~al.}(2018){H.~E.~S.~S.
  Collaboration}, {Abdalla}, {Abramowski}, {Aharonian}, {Ait Benkhali},
  {Akhperjanian}, {Ang{\"u}ner}, {Arrieta}, {Aubert}, {Backes}, {Balzer},
  {Barnard}, {Becherini}, {Becker Tjus}, {Berge}, {Bernhard}, {Bernl{\"o}hr},
  {Birsin}, {Blackwell}, {B{\"o}ttcher}, {Boisson}, {Bolmont}, {Bordas},
  {Bregeon}, {Brun}, {Brun}, {Bryan}, {Bulik}, {Capasso}, {Carr}, {Casanova},
  {Chakraborty}, {Chalme-Calvet}, {Chaves}, {Chen}, {Chevalier},
  {Chr{\'e}tien}, {Colafrancesco}, {Cologna}, {Condon}, {Conrad}, {Couturier},
  {Cui}, {Davids}, {Degrange}, {Deil}, {deWilt}, {Djannati-Ata{\"\i}},
  {Domainko}, {Donath}, {Drury}, {Dubus}, {Dutson}, {Dyks}, {Dyrda}, {Edwards},
  {Egberts}, {Eger}, {Ernenwein}, {Eschbach}, {Farnier}, {Fegan}, {Fernandes},
  {Fiasson}, {Fontaine}, {F{\"o}rster}, {Funk}, {F{\"u}{\ss}ling}, {Gabici},
  {Gajdus}, {Gallant}, {Garrigoux}, {Giavitto}, {Giebels}, {Glicenstein},
  {Gottschall}, {Goyal}, {Grondin}, {Grudzi{\'n}ska}, {Hadasch}, {Hahn},
  {Hawkes}, {Heinzelmann}, {Henri}, {Hermann}, {Hervet}, {Hillert}, {Hinton},
  {Hofmann}, {Hoischen}, {Holler}, {Horns}, {Ivascenko}, {Jacholkowska},
  {Jamrozy}, {Janiak}, {Jankowsky}, {Jankowsky}, {Jingo}, {Jogler}, {Jouvin},
  {Jung-Richardt}, {Kastendieck}, {Katarzy{\'n}ski}, {Katz}, {Kerszberg},
  {Kh{\'e}lifi}, {Kieffer}, {King}, {Klepser}, {Klochkov}, {Klu{\'z}niak},
  {Kolitzus}, {Komin}, {Kosack}, {Krakau}, {Kraus}, {Krayzel}, {Kr{\"u}ger},
  {Laffon}, {Lamanna}, {Lau}, {Lees}, {Lefaucheur}, {Lefranc}, {Lemi{\`e}re},
  {Lemoine-Goumard}, {Lenain}, {Leser}, {Lohse}, {Lorentz}, {Liu}, {Lypova},
  {Marandon}, {Marcowith}, {Mariaud}, {Marx}, {Maurin}, {Maxted}, {Mayer},
  {Meintjes}, {Menzler}, {Meyer}, {Mitchell}, {Moderski}, {Mohamed},
  {Mor{\r{a}}}, {Moulin}, {Murach}, {de Naurois}, {Niederwanger}, {Niemiec},
  {Oakes}, {Odaka}, {{\"O}ttl}, {Ohm}, {Ostrowski}, {Oya}, {Padovani},
  {Panter}, {Parsons}, {Paz Arribas}, {Pekeur}, {Pelletier}, {Petrucci},
  {Peyaud}, {Pita}, {Poon}, {Prokhorov}, {Prokoph}, {P{\"u}hlhofer}, {Punch},
  {Quirrenbach}, {Raab}, {Reimer}, {Reimer}, {Renaud}, {de Reyes}, {Rieger},
  {Romoli}, {Rosier-Lees}, {Rowell}, {Rudak}, {Rulten}, {Sahakian}, {Salek},
  {Sanchez}, {Santangelo}, {Sasaki}, {Schlickeiser}, {Sch{\"u}ssler}, {Schulz},
  {Schwanke}, {Schwemmer}, {Seyffert}, {Shafi}, {Shilon}, {Simoni}, {Sol},
  {Spanier}, {Spengler}, {Spies}, {Stawarz}, {Steenkamp}, {Stegmann},
  {Stinzing}, {Stycz}, {Sushch}, {Tavernet}, {Tavernier}, {Taylor}, {Terrier},
  {Tluczykont}, {Trichard}, {Tuffs}, {van der Walt}, {van Eldik}, {van Soelen},
  {Vasileiadis}, {Veh}, {Venter}, {Viana}, {Vincent}, {Vink}, {Voisin},
  {V{\"o}lk}, {Vuillaume}, {Wadiasingh}, {Wagner}, {Wagner}, {Wagner}, {White},
  {Wierzcholska}, {Willmann}, {W{\"o}rnlein}, {Wouters}, {Yang}, {Zabalza},
  {Zaborov}, {Zacharias}, {Zdziarski}, {Zech}, {Zefi}, {Ziegler}, \&
  {{\.Z}ywucka}}]{HESS-SGR1806:2018}
{H.~E.~S.~S. Collaboration}, {Abdalla}, H., {Abramowski}, A., {et~al.} 2018,
  \aap, 612, A11, \dodoi{10.1051/0004-6361/201628695}

\bibitem[{{H.E.S.S. Collaboration} {et~al.}(2017){H.E.S.S. Collaboration},
  {Abdalla}, {Abramowski}, {Aharonian}, {Ait Benkhali}, {Akhperjanian},
  {Andersson}, {Ang{\"u}ner}, {Arakawa}, {Arrieta}, {Aubert}, {Backes},
  {Balzer}, {Barnard}, {Becherini}, {Becker Tjus}, {Berge}, {Bernhard},
  {Bernl{\"o}hr}, {Blackwell}, {B{\"o}ttcher}, {Boisson}, {Bolmont}, {Bordas},
  {Bregeon}, {Brun}, {Brun}, {Bryan}, {B{\"u}chele}, {Bulik}, {Capasso},
  {Carr}, {Casanova}, {Cerruti}, {Chakraborty}, {Chalme-Calvet}, {Chaves},
  {Chen}, {Chevalier}, {Chr{\'e}tien}, {Coffaro}, {Colafrancesco}, {Cologna},
  {Condon}, {Conrad}, {Cui}, {Davids}, {Decock}, {Degrange}, {Deil}, {Devin},
  {Dewilt}, {Dirson}, {Djannati-Ata{\"\i}}, {Domainko}, {Donath}, {Drury},
  {Dutson}, {Dyks}, {Edwards}, {Egberts}, {Eger}, {Ernenwein}, {Eschbach},
  {Farnier}, {Fegan}, {Fernandes}, {Fiasson}, {Fontaine}, {F{\"o}rster},
  {Funk}, {F{\"u}{\ss}ling}, {Gabici}, {Gajdus}, {Gallant}, {Garrigoux},
  {Giavitto}, {Giebels}, {Glicenstein}, {Gottschall}, {Goyal}, {Grondin},
  {Hahn}, {Haupt}, {Hawkes}, {Heinzelmann}, {Henri}, {Hermann}, {Hervet},
  {Hinton}, {Hofmann}, {Hoischen}, {Holler}, {Horns}, {Ivascenko}, {Iwasaki},
  {Jacholkowska}, {Jamrozy}, {Janiak}, {Jankowsky}, {Jankowsky}, {Jingo},
  {Jogler}, {Jouvin}, {Jung-Richardt}, {Kastendieck}, {Katarzy{\'n}ski},
  {Katsuragawa}, {Katz}, {Kerszberg}, {Khangulyan}, {Kh{\'e}lifi}, {Kieffer},
  {King}, {Klepser}, {Klochkov}, {Klu{\'z}niak}, {Kolitzus}, {Komin}, {Kosack},
  {Krakau}, {Kraus}, {Kr{\"u}ger}, {Laffon}, {Lamanna}, {Lau}, {Lees},
  {Lefaucheur}, {Lefranc}, {Lemi{\`e}re}, {Lemoine-Goumard}, {Lenain}, {Leser},
  {Lohse}, {Lorentz}, {Liu}, {L{\'o}pez-Coto}, {Lypova}, {Marandon},
  {Marcowith}, {Mariaud}, {Marx}, {Maurin}, {Maxted}, {Mayer}, {Meintjes},
  {Meyer}, {Mitchell}, {Moderski}, {Mohamed}, {Mohrmann}, {Mor{\r{a}}},
  {Moulin}, {Murach}, {Nakashima}, {de Naurois}, {Niederwanger}, {Niemiec},
  {Oakes}, {O'Brien}, {Odaka}, {{\"O}ttl}, {Ohm}, {Ostrowski}, {Oya},
  {Padovani}, {Panter}, {Parsons}, {Pekeur}, {Pelletier}, {Perennes},
  {Petrucci}, {Peyaud}, {Piel}, {Pita}, {Poon}, {Prokhorov}, {Prokoph},
  {P{\"u}hlhofer}, {Punch}, {Quirrenbach}, {Raab}, {Reimer}, {Reimer},
  {Renaud}, {de Los Reyes}, {Richter}, {Rieger}, {Romoli}, {Rowell}, {Rudak},
  {Rulten}, {Sahakian}, {Saito}, {Salek}, {Sanchez}, {Santangelo}, {Sasaki},
  {Schlickeiser}, {Sch{\"u}ssler}, {Schulz}, {Schwanke}, {Schwemmer},
  {Seglar-Arroyo}, {Settimo}, {Seyffert}, {Shafi}, {Shilon}, {Simoni}, {Sol},
  {Spanier}, {Spengler}, {Spies}, {Stawarz}, {Steenkamp}, {Stegmann}, {Stycz},
  {Sushch}, {Takahashi}, {Tavernet}, {Tavernier}, {Taylor}, {Terrier},
  {Tibaldo}, {Tiziani}, {Tluczykont}, {Trichard}, {Tsuji}, {Tuffs}, {Uchiyama},
  {van der Walt}, {van Eldik}, {van Rensburg}, {van Soelen}, {Vasileiadis},
  {Veh}, {Venter}, {Viana}, {Vincent}, {Vink}, {Voisin}, {V{\"o}lk},
  {Vuillaume}, {Wadiasingh}, {Wagner}, {Wagner}, {Wagner}, {White},
  {Wierzcholska}, {Willmann}, {W{\"o}rnlein}, {Wouters}, {Yang}, {Zabalza},
  {Zaborov}, {Zacharias}, {Zanin}, {Zdziarski}, {Zech}, {Zefi}, {Ziegler},
  {{\.Z}ywucka}, {Superb Collaboration}, {Jankowski}, {Keane}, \&
  {Petroff}}]{2017A&A...597A.115H}
{H.E.S.S. Collaboration}, {Abdalla}, H., {Abramowski}, A., {et~al.} 2017, \aap,
  597, A115, \dodoi{10.1051/0004-6361/201629117}

\bibitem[{Holder \& Lynch(2019)}]{2019ICRC...36..698H}
Holder, J., \& Lynch, R.~S. 2019, Proc. ICRC 36, 698,
  \dodoi{10.22323/1.358.0698}

\bibitem[{Hu {et~al.}(2019)Hu, Baring, Wadiasingh, \& Harding}]{Hu_2019}
Hu, K., Baring, M.~G., Wadiasingh, Z., \& Harding, A.~K. 2019, \mnras, 486,
  3327–3349, \dodoi{10.1093/mnras/stz995}

\bibitem[{{Inan} {et~al.}(2007){Inan}, {Lehtinen}, {Moore}, {Hurley}, {Boggs},
  {Smith}, \& {Fishman}}]{Inan:2007}
{Inan}, U.~S., {Lehtinen}, N.~G., {Moore}, R.~C., {et~al.} 2007, \grl, 34,
  L08103, \dodoi{10.1029/2006GL029145}

\bibitem[{{Insight-HXMT}(2020)}]{HXMT_Bursts}
{Insight-HXMT}. 2020, SGR J1935+2154 burst list.
\newblock \url{http://hxmten.ihep.ac.cn/bfy/331.jhtml}

\bibitem[{{Ioka} {et~al.}(2005){Ioka}, {Razzaque}, {Kobayashi}, \&
  {M{\'e}sz{\'a}ros}}]{Ioka:2005}
{Ioka}, K., {Razzaque}, S., {Kobayashi}, S., \& {M{\'e}sz{\'a}ros}, P. 2005,
  \apj, 633, 1013, \dodoi{10.1086/466514}

\bibitem[{Kaneko {et~al.}(2014)Kaneko, Gogus, Younes, Guiriec, Kouveliotou, \&
  von Kienlin~A.}]{SGR1935_GBM1}
Kaneko, Y., Gogus, E., Younes, G., {et~al.} 2014, GCN 16577.
\newblock \url{https://gcn.gsfc.nasa.gov/gcn3/16577.gcn3}

\bibitem[{Kaspi \& Beloborodov(2017)}]{Kaspi:2017fwg}
Kaspi, V.~M., \& Beloborodov, A. 2017, ARA\&A, 55, 261,
  \dodoi{10.1146/annurev-astro-081915-023329}

\bibitem[{{Kirsten} {et~al.}(2021){Kirsten}, {Snelders}, {Jenkins}, {Nimmo},
  {van den Eijnden}, {Hessels}, {Gawro{\'n}ski}, \& {Yang}}]{Kirsten_2020}
{Kirsten}, F., {Snelders}, M.~P., {Jenkins}, M., {et~al.} 2021, NatAs, 5, 414,
  \dodoi{10.1038/s41550-020-01246-3}

\bibitem[{Kozlova {et~al.}(2016)Kozlova, Golenetskii, Aptekar, Frederiks,
  Oleynik, Ulanov, Svinkin, Tsvetkova, Lysenko, \&
  Cline}]{SGR1935_KONUS-WIND12}
Kozlova, A., Golenetskii, S., Aptekar, R., {et~al.} 2016, GCN 19438.
\newblock \url{https://gcn.gsfc.nasa.gov/gcn3/19438.gcn3}

\bibitem[{{Kozlova} {et~al.}(2016){Kozlova}, {Israel}, {Svinkin}, {Frederiks},
  {Pal'shin}, {Tsvetkova}, {Hurley}, {Goldsten}, {Golovin}, {Mitrofanov}, \&
  {Zhang}}]{SGR1935_KONUS-WIND10}
{Kozlova}, A.~V., {Israel}, G.~L., {Svinkin}, D.~S., {et~al.} 2016, \mnras,
  460, 2008, \dodoi{10.1093/mnras/stw1109}

\bibitem[{{Li} {et~al.}(2021){Li}, {Lin}, {Xiong}, {Ge}, {Li}, {Li}, {Lu},
  {Zhang}, {Tuo}, {Nang}, {Zhang}, {Xiao}, {Chen}, {Song}, {Xu}, {Liu}, {Jia},
  {Cao}, {Qu}, {Zhang}, {Gu}, {Liao}, {Zhao}, {Tan}, {Nie}, {Zhao}, {Zheng},
  {Zheng}, {Luo}, {Cai}, {Li}, {Xue}, {Bu}, {Chang}, {Chen}, {Chen}, {Chen},
  {Chen}, {Chen}, {Cui}, {Cui}, {Deng}, {Dong}, {Du}, {Fu}, {Gao}, {Gao},
  {Gao}, {Gu}, {Guan}, {Guo}, {Han}, {Huang}, {Huo}, {Jiang}, {Jiang}, {Jin},
  {Jin}, {Kong}, {Li}, {Li}, {Li}, {Li}, {Li}, {Li}, {Li}, {Liang}, {Liu},
  {Liu}, {Liu}, {Liu}, {Liu}, {Lu}, {Lu}, {Luo}, {Ma}, {Meng}, {Ou}, {Sai},
  {Shang}, {Song}, {Sun}, {Tao}, {Wang}, {Wang}, {Wang}, {Wang}, {Wang}, {Wen},
  {Wu}, {Wu}, {Wu}, {Xiao}, {Xu}, {Yang}, {Yang}, {Yang}, {Yang}, {Yi}, {Yin},
  {You}, {Zhang}, {Zhang}, {Zhang}, {Zhang}, {Zhang}, {Zhang}, {Zhang},
  {Zhang}, {Zhang}, {Zhang}, {Zhang}, {Zhang}, {Zhang}, {Zhang}, {Zhang},
  {Zhang}, {Zhou}, {Zhou}, {Zhu}, {Zhu}, \& {Zhuang}}]{li2020insighthxmt}
{Li}, C.~K., {Lin}, L., {Xiong}, S.~L., {et~al.} 2021, NatAs, 5, 378,
  \dodoi{10.1038/s41550-021-01302-6}

\bibitem[{Li \& Ma(1983)}]{LiMa}
Li, T.-P., \& Ma, Y. 1983, \apj, 272, 317, \dodoi{10.1086/161295}

\bibitem[{{Lin} {et~al.}(2020a){Lin}, {G{\"o}{\u{g}}{\"u}{\c{s}}}, {Roberts},
  {Kouveliotou}, {Kaneko}, {van der Horst}, \& {Younes}}]{Lin:2020-I}
{Lin}, L., {G{\"o}{\u{g}}{\"u}{\c{s}}}, E., {Roberts}, O.~J., {et~al.} 2020a,
  \apj, 893, 156, \dodoi{10.3847/1538-4357/ab818f}

\bibitem[{Lin {et~al.}(2020b)Lin, Gö{\u{g}}ü{\c{s}}, Roberts, Baring,
  Kouveliotou, Kaneko, van~der Horst, \& Younes}]{Lin_2020}
Lin, L., Gö{\u{g}}ü{\c{s}}, E., Roberts, O.~J., {et~al.} 2020b, \apjl, 902,
  L43, \dodoi{10.3847/2041-8213/abbefe}

\bibitem[{{Lin} {et~al.}(2020c){Lin}, {Zhang}, {Wang}, {Gao}, {Guan}, {Han},
  {Jiang}, {Jiang}, {Lee}, {Li}, {Men}, {Miao}, {Niu}, {Niu}, {Sun}, {Wang},
  {Wang}, {Xu}, {Xu}, {Xu}, {Yang}, {Yang}, {Yu}, {Zhang}, {Zhang}, {Zhou},
  {Zhu}, {Castro-Tirado}, {Dai}, {Ge}, {Hu}, {Li}, {Li}, {Li}, {Liang}, {Jia},
  {Querel}, {Shao}, {Wang}, {Wang}, {Wu}, {Xiong}, {Xu}, {Yang}, {Zhang},
  {Zhang}, {Zheng}, \& {Zou}}]{2020Natur.587...63L}
{Lin}, L., {Zhang}, C.~F., {Wang}, P., {et~al.} 2020c, Nature, 587, 63,
  \dodoi{10.1038/s41586-020-2839-y}

\bibitem[{Lyubarsky(2014)}]{Lyubarsky:2014jta}
Lyubarsky, Y. 2014, \mnras, 442, 9, \dodoi{10.1093/mnrasl/slu046}

\bibitem[{Mereghetti {et~al.}(2021)Mereghetti, Savchenko, Ferrigno, Bozzo,
  Gotz, L., \& Borkowski}]{INTEGRAL_2021}
Mereghetti, S., Savchenko, V., Ferrigno, C., {et~al.} 2021, GCN 29381.
\newblock \url{https://gcn.gsfc.nasa.gov/gcn3/29381.gcn3}

\bibitem[{{Mereghetti} {et~al.}(2020){Mereghetti}, {Savchenko}, {Ferrigno},
  {G{\"o}tz}, {Rigoselli}, {Tiengo}, {Bazzano}, {Bozzo}, {Coleiro},
  {Courvoisier}, {Doyle}, {Goldwurm}, {Hanlon}, {Jourdain}, {von Kienlin},
  {Lutovinov}, {Martin-Carrillo}, {Molkov}, {Natalucci}, {Onori}, {Panessa},
  {Rodi}, {Rodriguez}, {S{\'a}nchez-Fern{\'a}ndez}, {Sunyaev}, \&
  {Ubertini}}]{INTEGRAL_BURST_A}
{Mereghetti}, S., {Savchenko}, V., {Ferrigno}, C., {et~al.} 2020, \apjl, 898,
  L29, \dodoi{10.3847/2041-8213/aba2cf}

\bibitem[{{Metzger} {et~al.}(2019){Metzger}, {Margalit}, \&
  {Sironi}}]{Metzger:2019}
{Metzger}, B.~D., {Margalit}, B., \& {Sironi}, L. 2019, \mnras, 485, 4091,
  \dodoi{10.1093/mnras/stz700}

\bibitem[{{Parsons} \& {Hinton}(2014)}]{Parsons2014a}
{Parsons}, R.~D., \& {Hinton}, J.~A. 2014, APh, 56, 26,
  \dodoi{10.1016/j.astropartphys.2014.03.002}

\bibitem[{{Petroff} {et~al.}(2019){Petroff}, {Hessels}, \&
  {Lorimer}}]{Petroff:2019tty}
{Petroff}, E., {Hessels}, J.~W.~T., \& {Lorimer}, D.~R. 2019, \aapr, 27, 4,
  \dodoi{10.1007/s00159-019-0116-6}

\bibitem[{Petroff {et~al.}(2016)Petroff, Barr, Jameson, Keane, Bailes, Kramer,
  Morello, Tabbara, \& van Straten}]{Petroff:2016tcr}
Petroff, E., Barr, E.~D., Jameson, A., {et~al.} 2016, \pasa, 33, e045,
  \dodoi{10.1017/pasa.2016.35}

\bibitem[{{Petroff} {et~al.}(2017){Petroff}, {Burke-Spolaor}, {Keane},
  {McLaughlin}, {Miller}, {Andreoni}, {Bailes}, {Barr}, {Bernard}, {Bhandari},
  {Bhat}, {Burgay}, {Caleb}, {Champion}, {Chandra}, {Cooke}, {Dhillon},
  {Farnes}, {Hardy}, {Jaroenjittichai}, {Johnston}, {Kasliwal}, {Kramer},
  {Littlefair}, {Macquart}, {Mickaliger}, {Possenti}, {Pritchard}, {Ravi},
  {Rest}, {Rowlinson}, {Sawangwit}, {Stappers}, {Sullivan}, {Tiburzi}, {van
  Straten}, {ANTARES Collaboration}, {Albert}, {Andr{\'e}}, {Anghinolfi},
  {Anton}, {Ardid}, {Aubert}, {Avgitas}, {Baret}, {Barrios-Mart{\'\i}}, {Basa},
  {Bertin}, {Biagi}, {Bormuth}, {Bourret}, {Bouwhuis}, {Bruijn}, {Brunner},
  {Busto}, {Capone}, {Caramete}, {Carr}, {Celli}, {Chiarusi}, {Circella},
  {Coelho}, {Coleiro}, {Coniglione}, {Costantini}, {Coyle}, {Creusot},
  {Deschamps}, {de Bonis}, {Distefano}, {di Palma}, {Donzaud}, {Dornic},
  {Drouhin}, {Eberl}, {El Bojaddaini}, {Els{\"a}sser}, {Enzenh{\"o}fer},
  {Felis}, {Fusco}, {Galat{\`a}}, {Gay}, {Gei{\ss}els{\"o}der}, {Geyer},
  {Giordano}, {Gleixner}, {Glotin}, {Gr{\'e}goire}, {Gracia-Ruiz}, {Graf},
  {Hallmann}, {van Haren}, {Heijboer}, {Hello}, {Hern{\'a}ndez-Rey},
  {H{\"o}{\ss}l}, {Hofest{\"a}dt}, {Hugon}, {Illuminati}, {James}, {de Jong},
  {Jongen}, {Kadler}, {Kalekin}, {Katz}, {Kie{\ss}ling}, {Kouchner}, {Kreter},
  {Kreykenbohm}, {Kulikovskiy}, {Lachaud}, {Lahmann}, {Lef{\`e}vre}, {Leonora},
  {Lotze}, {Loucatos}, {Marcelin}, {Margiotta}, {Marinelli},
  {Mart{\'\i}nez-Mora}, {Mathieu}, {Mele}, {Melis}, {Michael}, {Migliozzi},
  {Moussa}, {Mueller}, {Nezri}, {P{\v{a}}v{\v{a}}la{\c{s}}}, {Pellegrino},
  {Perrina}, {Piattelli}, {Popa}, {Pradier}, {Quinn}, {Racca}, {Riccobene},
  {Roensch}, {S{\'a}nchez-Losa}, {Salda{\~n}a}, {Salvadori}, {Samtleben},
  {Sanguineti}, {Sapienza}, {Schnabel}, {Seitz}, {Sieger}, {Spurio},
  {Stolarczyk}, {Taiuti}, {Tayalati}, {Trovato}, {Tselengidou}, {Turpin},
  {T{\"o}nnis}, {Vallage}, {Vall{\'e}e}, {van Elewyck}, {Vivolo}, {Vizzoca},
  {Wagner}, {Wilms}, {Zornoza}, {Z{\'u}{\~n}iga}, {H.~E.~S.~S. Collaboration},
  {Abdalla}, {Abramowski}, {Aharonian}, {Ait Benkhali}, {Akhperjanian},
  {Andersson}, {Ang{\"u}ner}, {Arrieta}, {Aubert}, {Backes}, {Balzer},
  {Barnard}, {Becherini}, {Tjus}, {Berge}, {Bernhard}, {Bernl{\"o}hr},
  {Blackwell}, {B{\"o}ttcher}, {Boisson}, {Bolmont}, {Bordas}, {Bregeon},
  {Brun}, {Brun}, {Bryan}, {Bulik}, {Capasso}, {Casanova}, {Cerruti},
  {Chakraborty}, {Chalme-Calvet}, {Chaves}, {Chen}, {Chevalier},
  {Chr{\'e}tien}, {Colafrancesco}, {Cologna}, {Condon}, {Conrad}, {Cui},
  {Davids}, {Decock}, {Degrange}, {Deil}, {Devin}, {Dewilt}, {Dirson},
  {Djannati-Ata{\"\i}}, {Domainko}, {Donath}, {Drury}, {Dubus}, {Dutson},
  {Dyks}, {Edwards}, {Egberts}, {Eger}, {Ernenwein}, {Eschbach}, {Farnier},
  {Fegan}, {Fernandes}, {Fiasson}, {Fontaine}, {F{\"o}rster}, {Funk},
  {F{\"u}{\ss}ling}, {Gabici}, {Gajdus}, {Gallant}, {Garrigoux}, {Giavitto},
  {Giebels}, {Glicenstein}, {Gottschall}, {Goyal}, {Grondin}, {Hadasch},
  {Hahn}, {Haupt}, {Hawkes}, {Heinzelmann}, {Henri}, {Hermann}, {Hervet},
  {Hinton}, {Hofmann}, {Hoischen}, {Holler}, {Horns}, {Ivascenko},
  {Jacholkowska}, {Jamrozy}, {Janiak}, {Jankowsky}, {Jankowsky}, {Jingo},
  {Jogler}, {Jouvin}, {Jung-Richardt}, {Kastendieck}, {Katarzy{\'n}ski},
  {Kerszberg}, {Kh{\'e}lifi}, {Kieffer}, {King}, {Klepser}, {Klochkov},
  {Klu{\'z}niak}, {Kolitzus}, {Komin}, {Kosack}, {Krakau}, {Kraus}, {Krayzel},
  {Kr{\"u}ger}, {Laffon}, {Lamanna}, {Lau}, {Lees}, {Lefaucheur}, {Lefranc},
  {Lemi{\`e}re}, {Lemoine-Goumard}, {Lenain}, {Leser}, {Lohse}, {Lorentz},
  {Liu}, {L{\'o}pez-Coto}, {Lypova}, {Marandon}, {Marcowith}, {Mariaud},
  {Marx}, {Maurin}, {Maxted}, {Mayer}, {Meintjes}, {Meyer}, {Mitchell},
  {Moderski}, {Mohamed}, {Mohrmann}, {Mor{\^a}}, {Moulin}, {Murach}, {de
  Naurois}, {Niederwanger}, {Niemiec}, {Oakes}, {O'Brien}, {Odaka}, {{\"O}ttl},
  {Ohm}, {Ostrowski}, {Oya}, {Padovani}, {Panter}, {Parsons}, {Pekeur},
  {Pelletier}, {Perennes}, {Petrucci}, {Peyaud}, {Piel}, {Pita}, {Poon},
  {Prokhorov}, {Prokoph}, {P{\"u}hlhofer}, {Punch}, {Quirrenbach}, {Raab},
  {Reimer}, {Reimer}, {Renaud}, {Reyes}, {Rieger}, {Romoli}, {Rosier-Lees},
  {Rowell}, {Rudak}, {Rulten}, {Sahakian}, {Salek}, {Sanchez}, {Santangelo},
  {Sasaki}, {Schlickeiser}, {Schulz}, {Sch{\"u}ssler}, {Schwanke}, {Schwemmer},
  {Settimo}, {Seyffert}, {Shafi}, {Shilon}, {Simoni}, {Sol}, {Spanier},
  {Spengler}, {Spies}, {Stawarz}, {Steenkamp}, {Stegmann}, {Stinzing}, {Stycz},
  {Sushch}, {Tavernet}, {Tavernier}, {Taylor}, {Terrier}, {Tibaldo}, {Tiziani},
  {Tluczykont}, {Trichard}, {Tuffs}, {Uchiyama}, {Walt}, {van Eldik}, {van
  Rensburg}, {van Soelen}, {Vasileiadis}, {Veh}, {Venter}, {Viana}, {Vincent},
  {Vink}, {Voisin}, {V{\"o}lk}, {Vuillaume}, {Wadiasingh}, {Wagner}, {Wagner},
  {Wagner}, {White}, {Wierzcholska}, {Willmann}, {W{\"o}rnlein}, {Wouters},
  {Yang}, {Zabalza}, {Zaborov}, {Zacharias}, {Zanin}, {Zdziarski}, {Zech},
  {Zefi}, {Ziegler}, \& {{\.Z}ywucka}}]{2017MNRAS.469.4465P}
{Petroff}, E., {Burke-Spolaor}, S., {Keane}, E.~F., {et~al.} 2017, \mnras, 469,
  4465, \dodoi{10.1093/mnras/stx1098}

\bibitem[{{Popov} \& {Postnov}(2010)}]{Popov:2007uv}
{Popov}, S.~B., \& {Postnov}, K.~A. 2010, in Evolution of Cosmic Objects
  through their Physical Activity, ed. H.~A. {Harutyunian}, A.~M. {Mickaelian},
  \& Y.~{Terzian}, 129--132.
\newblock \doarXiv{0710.2006}

\bibitem[{Ricciarini {et~al.}(2021)Ricciarini, Yoshida, Sakamoto, Pal'shin,
  Sugita, Kawakubo, Yamaoka, Nakahira, Asaoka, Torii, Akaike, Kobayashi,
  Shimizu, Tamura, Cannady, Cherry, \& Marrocchesi}]{CGBM_2021}
Ricciarini, S., Yoshida, A., Sakamoto, T., {et~al.} 2021, GCN 29383.
\newblock \url{https://gcn.gsfc.nasa.gov/gcn3/29383.gcn3}

\bibitem[{Ridnaia {et~al.}(2019)Ridnaia, Golenetskii, Aptekar, Frederiks,
  Ulanov, Svinkin, Tsvetkova, Lysenko, \& Cline}]{SGR1935_KONUS-WIND7}
Ridnaia, A., Golenetskii, S., Aptekar, R., {et~al.} 2019, GCN 26242.
\newblock \url{https://gcn.gsfc.nasa.gov/gcn3/26242.gcn3}

\bibitem[{Ridnaia {et~al.}(2020{\natexlab{a}})Ridnaia, Golenetskii, Aptekar,
  Frederiks, Ulanov, Svinkin, Tsvetkova, Lysenko, \&
  Cline}]{SGR1935_KONUS-WIND2}
---. 2020{\natexlab{a}}, ATel, 13688.
\newblock \url{http://www.astronomerstelegram.org/?read=13688}

\bibitem[{Ridnaia {et~al.}(2020{\natexlab{b}})Ridnaia, Golenetskii, Aptekar,
  Frederiks, Ulanov, Svinkin, Tsvetkova, Lysenko, \&
  Cline}]{SGR1935_KONUS-WIND3}
---. 2020{\natexlab{b}}, GCN 27715.
\newblock \url{https://gcn.gsfc.nasa.gov/gcn/gcn3/27715.gcn3}

\bibitem[{Ridnaia {et~al.}(2020{\natexlab{c}})Ridnaia, Golenetskii, Aptekar,
  Frederiks, Ulanov, Svinkin, Tsvetkova, Lysenko, \&
  Cline}]{SGR1935_KONUS-WIND4}
---. 2020{\natexlab{c}}, GCN 27554.
\newblock \url{https://gcn.gsfc.nasa.gov/gcn/gcn3/27554.gcn3}

\bibitem[{Ridnaia {et~al.}(2020{\natexlab{d}})Ridnaia, Golenetskii, Aptekar,
  Frederiks, Ulanov, Svinkin, Tsvetkova, Lysenko, \&
  Cline}]{SGR1935_KONUS-WIND6}
---. 2020{\natexlab{d}}, GCN 29373.
\newblock \url{https://gcn.gsfc.nasa.gov/gcn3/29373.gcn3}

\bibitem[{Ridnaia {et~al.}(2020{\natexlab{e}})Ridnaia, Golenetskii, Aptekar,
  Frederiks, Ulanov, Svinkin, Tsvetkova, Lysenko, \&
  Cline}]{SGR1935_KONUS-WIND8}
---. 2020{\natexlab{e}}, GCN 27631.
\newblock \url{https://gcn.gsfc.nasa.gov/gcn3/27631.gcn3}

\bibitem[{Ridnaia {et~al.}(2020{\natexlab{f}})Ridnaia, Golenetskii, Aptekar,
  Frederiks, Ulanov, Svinkin, Tsvetkova, Lysenko, \&
  Cline}]{SGR1935_KONUS-WIND9}
---. 2020{\natexlab{f}}, GCN 27667.
\newblock \url{https://gcn.gsfc.nasa.gov/gcn3/27667.gcn3}

\bibitem[{{Ridnaia} {et~al.}(2021){Ridnaia}, {Svinkin}, {Frederiks}, {Bykov},
  {Popov}, {Aptekar}, {Golenetskii}, {Lysenko}, {Tsvetkova}, {Ulanov}, \&
  {Cline}}]{Ridnaia:2021}
{Ridnaia}, A., {Svinkin}, D., {Frederiks}, D., {et~al.} 2021, NatAs, 5, 372,
  \dodoi{10.1038/s41550-020-01265-0}

\bibitem[{Ridnaia {et~al.}(2021)Ridnaia, Golenetskii, Aptekar, Frederiks,
  Ulanov, Svinkin, Tsvetkova, Lysenko, \& on~behalf of~the
  Konus-Wind~team}]{SGR1935_KONUS-WIND_2021}
Ridnaia, A., Golenetskii, S., Aptekar, R., {et~al.} 2021, GCN 29373.
\newblock \url{https://gcn.gsfc.nasa.gov/gcn/gcn3/29373.gcn3}

\bibitem[{Roberts {et~al.}(2021{\natexlab{a}})Roberts, Wood, von Kienlin,
  Veres, \& Younes}]{SGR1935_GBM_2021-1}
Roberts, O., Wood, J., von Kienlin, A., Veres, P., \& Younes, G.
  2021{\natexlab{a}}, GCN 29374.
\newblock \url{https://gcn.gsfc.nasa.gov/gcn3/29374.gcn3}

\bibitem[{Roberts {et~al.}(2021{\natexlab{b}})Roberts, Wood, von Kienlin,
  Veres, \& Younes}]{SGR1935_GBM_2021-2}
---. 2021{\natexlab{b}}, GCN 29388.
\newblock \url{https://gcn.gsfc.nasa.gov/gcn3/29388.gcn3}

\bibitem[{{Rolke} {et~al.}(2005){Rolke}, {L{\'o}pez}, \& {Conrad}}]{Rolke}
{Rolke}, W.~A., {L{\'o}pez}, A.~M., \& {Conrad}, J. 2005, NIMPA, 551, 493,
  \dodoi{10.1016/j.nima.2005.05.068}

\bibitem[{Scholz(2020)}]{SGR1935_CHIME2}
Scholz, P. 2020, ATel, 13681.
\newblock \url{http://www.astronomerstelegram.org/?read=13681}

\bibitem[{{Tavani} {et~al.}(2021){Tavani}, {Casentini}, {Ursi}, {Verrecchia},
  {Addis}, {Antonelli}, {Argan}, {Barbiellini}, {Baroncelli}, {Bernardi},
  {Bianchi}, {Bulgarelli}, {Caraveo}, {Cardillo}, {Cattaneo}, {Chen}, {Costa},
  {Del Monte}, {Di Cocco}, {Di Persio}, {Donnarumma}, {Evangelista}, {Feroci},
  {Ferrari}, {Fioretti}, {Fuschino}, {Galli}, {Gianotti}, {Giuliani},
  {Labanti}, {Lazzarotto}, {Lipari}, {Longo}, {Lucarelli}, {Magro},
  {Marisaldi}, {Mereghetti}, {Morelli}, {Morselli}, {Naldi}, {Pacciani},
  {Parmiggiani}, {Paoletti}, {Pellizzoni}, {Perri}, {Perotti}, {Piano},
  {Picozza}, {Pilia}, {Pittori}, {Puccetti}, {Pupillo}, {Rapisarda},
  {Rappoldi}, {Rubini}, {Setti}, {Soffitta}, {Trifoglio}, {Trois},
  {Vercellone}, {Vittorini}, {Giommi}, \& {D'Amico}}]{tavani2020xray}
{Tavani}, M., {Casentini}, C., {Ursi}, A., {et~al.} 2021, NatAs, 5, 401,
  \dodoi{10.1038/s41550-020-01276-x}

\bibitem[{Thompson \& Duncan(2001)}]{Thompson_2001}
Thompson, C., \& Duncan, R.~C. 2001, \apj, 561, 980–1005,
  \dodoi{10.1086/323256}

\bibitem[{Tohuvavohu(2020)}]{Swift-BAT_ATel}
Tohuvavohu, A. 2020, ATel, 13758.
\newblock \url{http://www.astronomerstelegram.org/?read=13758}

\bibitem[{Ursi {et~al.}(2020)Ursi, Pittori, Tempesta, Verrecchia, Tavani,
  Cardillo, Casentini, Piano, {et~al.}}]{AGILE_ATEL}
Ursi, A., Pittori, C., Tempesta, P., {et~al.} 2020, ATel, 13682.
\newblock \url{http://www.astronomerstelegram.org/?read=13682}

\bibitem[{von Kienlin {et~al.}(2014)von Kienlin, Meegan, Paciesas, Bhat,
  Bissaldi, Briggs, Burgess, Byrne, Chaplin, Cleveland, Connaughton, Collazzi,
  Fitzpatrick, Foley, Gibby, Giles, Goldstein, Greiner, Gruber, Guiriec,
  van~der Horst, Kouveliotou, Layden, McBreen, McGlynn, Pelassa, Preece, Rau,
  Tierney, Wilson-Hodge, Xiong, Younes, \& Yu}]{von_Kienlin_2014}
von Kienlin, A., Meegan, C.~A., Paciesas, W.~S., {et~al.} 2014, \apjs, 211, 13,
  \dodoi{10.1088/0067-0049/211/1/13}

\bibitem[{von Kienlin {et~al.}(2020)von Kienlin, Meegan, Paciesas, Bhat,
  Bissaldi, Briggs, Burns, Cleveland, Gibby, Giles, Goldstein, Hamburg, Hui,
  Kocevski, Mailyan, Malacaria, Poolakkil, Preece, Roberts, Veres, \&
  Wilson-Hodge}]{von_Kienlin_2020}
---. 2020, \apj, 893, 46, \dodoi{10.3847/1538-4357/ab7a18}

\bibitem[{Younes \& Kouveliotou(2016)}]{SGR1935_GBM5}
Younes, G., \& Kouveliotou, C. 2016, GCN 19546.
\newblock \url{https://gcn.gsfc.nasa.gov/gcn3/19546.gcn3}

\bibitem[{Younes {et~al.}(2016)Younes, Kouveliotou, Hamburg, \&
  Burns}]{SGR1935_GBM6}
Younes, G., Kouveliotou, C., Hamburg, R., \& Burns, E. 2016, GCN 19598.
\newblock \url{https://gcn.gsfc.nasa.gov/gcn3/19598.gcn3}

\bibitem[{Younes(2016)}]{SGR1935_GBM4}
Younes, G.~Y. 2016, GCN 19437.
\newblock \url{https://gcn.gsfc.nasa.gov/gcn3/19437.gcn3}

\bibitem[{Yu \& Veres(2016)}]{SGR1935_GBM3}
Yu, H.-F., \& Veres, P. 2016, GCN 19434.
\newblock \url{https://gcn.gsfc.nasa.gov/gcn3/19434.gcn3}

\bibitem[{{Zhang}(2020)}]{2020Natur.587...45Z}
{Zhang}, B. 2020, \nat, 587, 45, \dodoi{10.1038/s41586-020-2828-1}

\bibitem[{Zhang {et~al.}(2020{\natexlab{a}})Zhang, Mereghetti, Savchenko,
  Ferrigno, Götz, Rigoselli, Tiengo, Bazzano, Bozzo, Coleiro, Courvoisier, \&
  et~al.}]{FAST_FRB}
Zhang, C.~F., Mereghetti, S., Savchenko, V., {et~al.} 2020{\natexlab{a}}, ATel,
  13699.
\newblock \url{http://www.astronomerstelegram.org/?read=13699}

\bibitem[{Zhang {et~al.}(2020{\natexlab{b}})Zhang, Tuo, Xiong, C.-K, Xiao,
  S.-M., {et~al.}}]{HXMT_Bursts2}
Zhang, S.-N., Tuo, Y.-L., Xiong, S.-L., {et~al.} 2020{\natexlab{b}}, ATel,
  13687.
\newblock \url{http://www.astronomerstelegram.org/?read=13687}

\bibitem[{Zhang {et~al.}(2020{\natexlab{c}})Zhang, Zhang, \& Lu}]{HXMT_Delay}
Zhang, S.-N., Zhang, B., \& Lu, W.-B. 2020{\natexlab{c}}, ATel, 13692.
\newblock \url{http://www.astronomerstelegram.org/?read=13692}

\bibitem[{{Zhou} {et~al.}(2020){Zhou}, {Zhou}, {Chen}, {Wang}, {Vink}, \&
  {Wang}}]{Zhou:2020}
{Zhou}, P., {Zhou}, X., {Chen}, Y., {et~al.} 2020, \apj, 905, 99,
  \dodoi{10.3847/1538-4357/abc34a}

\end{thebibliography}
